\documentclass[a4paper,11pt]{article}
\usepackage{jheppub} % for details on the use of the package, please see the JINST-author-manual
\usepackage{slashed}
\usepackage{xcolor}
\usepackage[normalem]{ulem} 

\def\i{{\mathsf i}}

\def\Hom{\mathop{\mathrm{Hom}}}

\def\cA{{\cal A}}

\def\cC{{\cal C}}
\def\cD{{\cal D}}

\def\cL{{\cal L}}
\def\cM{{\cal M}}
\def\cN{{\cal N}}

\def\bZ{{\mathbb Z}}

\def\sR{{\mathsf R}}
\def\sS{{\mathsf S}}
\def\sT{{\mathsf T}}

\def\sc{{\mathsf c}}

\def\sg{{\mathsf g}}
\def\sh{{\mathsf h}}

\def\U{\mathrm{U}}
\def\SU{\mathrm{SU}}

\def\Sp{\mathrm{Sp}}

\def\Pin{\mathrm{Pin}}
\def\SL{\mathrm{SL}}

\def\GL{\mathrm{GL}}

\def\g{\mathfrak{g}}

\title{{$\Sp(4,\mathbb{Z})$ actions on 3d $U(1)^2$ symmetric theories: Order-five duality and bilayer quantum Hall hierarchies}}

\author[a]{Yasin F. Alam,}
\author[a]{Andreas Karch,}
\author[b,c]{Da-Chuan Lu,}
\author[a]{Ryan C. Spieler}
\affiliation[a]{Theory Group, Weinberg Institute, Department of Physics,
University of Texas \\  2515 Speedway, Austin, TX 78712, USA.}
\affiliation[b]{Department of Physics, Harvard University, Cambridge, MA 02138, USA}
\affiliation[c]{Department of Physics and Center for Theory of Quantum Matter, University of Colorado, Boulder, CO, 80309, USA}
\emailAdd{yfralam@utexas.edu, karcha@utexas.edu, dclu137@gmail.com, rcspieler@utexas.edu}

\abstract{
The $\SL(2,\mathbb{Z})$ electromagnetic duality of 4d Maxwell theory induces theory-generating operations on 3d theories with $U(1)$ global symmetry. For theories with $U(1)^n$ symmetry, this structure generalizes to $\Sp(2n,\mathbb{Z})$. Focusing on the $U(1)^2$ case, we formulate the bulk $\Sp(4,\mathbb{Z})$ action and derive the corresponding boundary operations. We identify an intrinsically two-component element of order five, which generalizes the order-three $ST$ element of the single-$U(1)$ theory. Although its fifth power acts trivially in the bulk, the corresponding boundary operation closes only up to a decoupled $U(1)_1$ invertible phase, suggesting a mixed duality-gravitational anomaly. We realize the resulting theory-generating web for Abelian Chern-Simons theories within the $K$-matrix formalism and apply it to bilayer fractional quantum Hall systems. Recasting the Haldane--Halperin hierarchy construction as a sequence of $\SL(2,\mathbb{Z})$ operations, we generalize it to systems with charge $U(1)_c$ and pseudospin $U(1)_s$ symmetries. The resulting bilayer hierarchies contain branches terminating in an interlayer-correlated bosonic $(221)$ daughter sector, yielding candidate Abelian states at the even-denominator equal-layer fillings $3/8+3/8$ and $5/12+5/12$. Integral changes of anyon basis establish the equivalence of these states to their corresponding Abelian composite-fermion descriptions. We further discuss the spin-charge constraints that arise when the electromagnetic background field is treated as a spin-$c$ connection.}

\begin{document}
\maketitle
\flushbottom

\section{Introduction}

As first described in \cite{Witten:2003ya}, three dimensional conformal field theories with a global $U(1)$ symmetry allow for interesting transformations that systematically generate new CFTs from old ones by gauging the global symmetry, potentially in the presence of extra topological terms. These transformations also feature in the well studied 3d duality web \cite{Seiberg:2016gmd,Karch:2016sxi,Wang:2017txt}, where they play the role of theory-generating transformations. That is they do {\it not} by themselves correspond to dualities in the traditional high energy theory sense of the word in that they do not take one theory into a physically equivalent one; but they can be used to generate new dual pairs from old ones when the same transformation is applied to both sides of a known dual pair.  We note that later in the paper we use duality in a weaker sense often encountered in condensed matter literature where we do not demand that the operation we use takes a theory to a physically equivalent one.

The extension to theories with $U(1)^n$ global symmetry is governed by the symplectic electric--magnetic duality action of 4d Abelian gauge theory \cite{Gaillard:1981rj,Dimofte:2011ju}. Its induced action on 3d boundary theories can be formulated in terms of duality walls and boundary conditions \cite{Kapustin:2009av,Dimofte:2011ju}. 
The case $n=2$ is the simplest setting that goes beyond two independent copies of the single-$U(1)$ construction. In addition to transformations acting separately on the two $U(1)$ factors, the full $\Sp(4,\mathbb{Z})$ action contains mixed Chern--Simons or BF counterterms, gauging operations in mixed charge bases, and $\GL(2,\mathbb{Z})$ redefinitions of the symmetry currents \cite{Dimofte:2011ju,hua1949generators}. The $U(1)^2$ case is also special from the viewpoint of duality anomalies. The $\Sp(4,\mathbb{Z})$ action can be refined using the genus-two mapping class group $\Gamma_2$, for which the mixed duality--gravity anomaly has a finite $\mathbb{Z}_{10}$ classification \cite{Seiberg:2018ntt,KorkmazStipsicz2003}. This finite structure is particular to low rank: for $U(1)^n$ with $n\geq3$, the corresponding anomaly is instead characterized by an integer-valued class \cite{Harer1983,Seiberg:2018ntt}. The $U(1)^2$ theory therefore provides the simplest setting beyond a single $U(1)$ in which to study finite-order duality relations and their projective boundary realizations.

Another motivation is to understand the duality web of phases with $U(1)\times U(1)$ symmetry. In contrast to webs based on finite product symmetries \cite{Karch:2025hsj}, continuous symmetries introduce additional infrared subtleties. Their spontaneous breaking produces gapless Nambu--Goldstone modes \cite{Watanabe:2020ngm}, so the resulting phases need not admit a purely topological low-energy description. Symmetry-preserving gapped phases may instead be invertible phases characterized by quantized response \cite{Lu:2012dt}, or symmetry-enriched topological phases with intrinsic topological order and nontrivial symmetry-fractionalization data \cite{Lu:2013jqa,Barkeshli:2019vtb}. For Abelian topological phases, these data are encoded by a $K$-matrix and two charge vectors, which determine the separate and mixed $U(1)$ responses \cite{Wen:1995qn,Lu:2013jqa}. Gauging a continuous symmetry can further change the topological order in a manner controlled by its Hall response and fractionalization data \cite{Cheng:2022nds,Cheng:2022jbr}. The $\Sp(4,\mathbb{Z})$ duality web can therefore organize symmetry-breaking, invertible phases, and symmetry-enriched topological phases with two conserved $U(1)$ charges.

Motivated by these technical and physical considerations, we investigate the $\Sp(4,\mathbb{Z})$ action from two complementary perspectives. First, we study genuinely two-component duality relations and their projective realization on the boundary. Second, we use the $\Sp(4,\mathbb{Z})$ theory-generating operations to organize $U(1)\times U(1)$-symmetric phases and, in particular, bilayer fractional quantum Hall states.

A particularly interesting feature of the bulk--boundary construction is that the boundary operations can realize the duality group projectively. For a single $U(1)$, the order-three element $ST$ satisfies $(\sS\sT)^3=I$ in the bulk, whereas its boundary action closes only up to a decoupled $U(1)_1$ invertible phase. We find an intrinsically two-component analogue in $\Sp(4,\mathbb{Z})$: an order-five transformation $g_5$ whose bulk matrix satisfies $\sg_5^5=I$, while its boundary action obeys $g_5^5=U(1)_1$. This projective phase cannot be removed by redefining $g_5$ through stacking with an invertible phase and provides a boundary signature of a mixed duality-gravity anomaly. Its precise relation to the anomaly formulated through the genus-two mapping class group $\Gamma_2$ remains an open question \cite{Seiberg:2018ntt}.

Beyond this finite-order relation, the $\Sp(4,\mathbb{Z})$ operations provide a constructive framework for generating and organizing phases with two conserved $U(1)$ charges. The connection between modular transformations and fractional quantum Hall states was investigated before with different focuses \cite{Burgess:2000kj,Burgess:2001sy,Burgess:2007qa,Lutken:2011zz,NissinenLutken2012,GeraedtsMotrunich2013,Son:2015xqa,Seiberg:2016gmd,Karch:2016sxi,Goldman:2018zfm,Cheng:2022jbr}. To make this connection concrete, we first revisit the conventional Haldane--Halperin hierarchy \cite{Haldane:1983xm,Halperin:1984zz}. Tuning away from a parent fractional quantum Hall state creates quasiparticles or quasiholes, which may themselves form a daughter Laughlin state. We show that an elementary hierarchy step can be expressed as an $ST^p$ operation on the background field coupled to the excess-quasiparticle current: $T^p$ introduces the level-$p$ Chern--Simons term describing the daughter state, while $S$ promotes the corresponding background field to a dynamical gauge field. Repeated applications enlarge the $K$-matrix and reproduce the familiar continued-fraction structure of the hierarchy. The single-component hierarchy can therefore be organized as a sequence of $\SL(2,\mathbb{Z})$ theory-generating operations.

We then generalize this construction to bilayer systems with separately conserved layer charges, equivalently described by charge $U(1)_c$ and pseudospin $U(1)_s$ symmetries. Starting from a Halperin $(m,m,n)$ parent state \cite{Halperin:1983mmn}, the $\Sp(4,\mathbb{Z})$ operations act simultaneously on the quasiparticle currents of the two layers. Consequently, the daughter sector need not consist of two independent Laughlin states: the excess quasiparticles or quasiholes can instead form a correlated bilayer Halperin state, with the off-diagonal entries of its $K$-matrix encoding interlayer correlations. This construction generates layer-exchange-symmetric and $SU(2)$-compatible spin-singlet hierarchies, together with their associated continued-fraction sequences. In particular, starting from the $(330)$ parent, a final interlayer-correlated bosonic $(221)$ daughter sector yields candidate Abelian states at
\begin{equation}
    (330)+(221)\;:\quad \nu_1=\nu_2=\frac{3}{8},
    \qquad
    (330)+(220)+(221)\;:\quad \nu_1=\nu_2=\frac{5}{12}.
\end{equation}
These fractions are relevant to recent graphene-bilayer experiments \cite{Nguyen:2024jip}. We further show through explicit $\GL(N,\mathbb{Z})$ changes of anyon basis that the resulting hierarchy theories describe the same low-energy Abelian topological orders as the corresponding bilayer composite-fermion constructions \cite{Jain:1989zz,scarola2001phase}.

This paper is organized as follows. Section~\ref{sec:operations} reviews the \(\SL(2,\mathbb{Z})\) action for a single \(U(1)\), develops the \(\Sp(2n,\mathbb{Z})\) electromagnetic duality transformations and their induced boundary operations, and specializes to \(U(1)^2\), including the projective relations and the order-five element \(g_5\). Section~\ref{sec:kmatrix} formulates these operations for Abelian Chern-Simons theories in the \(K\)-matrix formalism and discusses the resulting \(U(1)\) and \(U(1)\times U(1)\) symmetric phases. Section~\ref{sec:bilayer} applies the construction to single-component and bilayer fractional quantum Hall hierarchies, including layer-exchange-symmetric and spin-singlet branches, their relation to composite-fermion descriptions, and the constraints from spin-charge relations. Section~\ref{sec:conclusion} summarizes our results and discusses future directions. The appendices collect a presentation of \(\Sp(4,\mathbb{Z})\), an extended matrix description of the theory-generating web, and details of the duality anomaly, the projective order-five relation and analogous $\bZ_2\times\bZ_2$ action.

\section{Bulk duality and boundary operations for $U(1)^n$ theories}
\label{sec:operations}
In this section, we review the $\Sp(2n,\bZ)$ electromagnetic-duality action of 4d $U(1)^n$ Maxwell theory and derive the corresponding duality operations on a 3d boundary. We begin with the $n=1$ case, which fixes our conventions for the $\SL(2,\bZ)$ generators and their boundary action. We then describe the bulk $\Sp(2n,\bZ)$ transformations and the induced $S$-, $T$-, and $\GL(n,\bZ)$-type boundary operations. Finally, we specialize to $n=2$, introduce a convenient generating set for $\Sp(4,\bZ)$, and study its boundary relations, including a novel order-five element that intrinsically mixes the two $U(1)$ factors with projective action.

We denote the bulk symplectic matrices by $\sS,\sT,\ldots$ and the corresponding boundary operations by $S,T,\ldots$. Products act from right to left. The bulk matrices obey the $\Sp(2n,\bZ)$ group relations exactly, whereas their boundary lifts may satisfy these relations only up to stacking with an invertible 3d phase.

\subsection{The $\SL(2,\bZ)$ action for a single $U(1)$}

It has long been known that there is a natural action of $\SL(2,\bZ)$ on 3d quantum field theories (QFTs) with $\U(1)$ symmetry group \cite{Witten:2003ya}. A natural way to obtain the action of the duality group, $\SL(2,\bZ)$, on the 3d theory, is to consider 4d $\U(1)$ Maxwell theory coupled to a 3d theory at the boundary, and then consider the action of the generators of the duality group \cite{Seiberg:2016gmd}. 

Given a theory with partition function dependent on some background gauge field $A$ \footnote{Throughout this paper, we write gauge fields and their field strengths using differential forms.  This is sufficient, since we only consider manifolds without homologically nontrivial cycles.  We suppress the wedge product between differential forms.}, $Z[A]$, the two generators of $\SL(2,\bZ)$, $S$ and $T$ respectively, act by gauging the $U(1)$ symmetry \footnote{It might be worth stressing that we do not restrict ourselves to the topological operation of summing over only flat gauge fields - our gauge fields are allowed nonvanishing field strengths.  For recent discussions of topological and nontopological operations involving continuous symmetries and their connection to field theories in one dimension higher, see i.e. \cite{Brennan:2024fgj, Antinucci:2024zjp, Apruzzi:2024htg, Jia:2026tfh}.} 
\begin{align}
    S: Z[A] \to Z_S[A]=\int \cD a \exp{\left(\frac{\i}{2\pi} \int_{Y_3}adA\right)}Z[a] \, ,
\end{align}
and stacking a level one $\U(1)$ background Chern-Simons term
\begin{align}
    T:Z[A] \to Z_T[A]=\exp{\left(\frac{\i}{4\pi} \int_{Y_3}A dA\right)} Z[A] \, ,
\end{align}
where lower case $a$ denotes a dynamical field. It is not difficult to check that $S^2$ acts as charge conjugation ($Z_{S^2}[A]=Z[-A])$. There is an order-3 action $ST$ which corresponds to the twisted gauging, and $(ST)^3$ acts by stacking a dynamical level one Chern-Simons theory
\begin{align}
    Z_{(ST)^3}[A] =  \int \cD a \exp{\left(\frac{\i}{4\pi} \int_{Y_3}a da\right)}Z[A] \, .
    \label{eq:proj-rep}
\end{align}
Interestingly, $S$ and $T$ only form a projective representation of $\SL(2,\bZ)$ due to the $(ST)^3$ failing to give the identity. Indeed, one \textit{cannot} redefine $S$ and $T$ in order to make a linear $\SL(2,\bZ)$ action \cite{Seiberg:2018ntt,Bhardwaj:2020ymp}.

\subsection{$\Sp(2n,\bZ)$ electromagnetic duality in four dimensions}
We now generalize the construction to $n$ Abelian gauge fields. The electromagnetic-duality group of 4d $U(1)^n$ Maxwell theory is $\Sp(2n,\bZ)$, which acts on the complexified coupling matrix and mixes electric and magnetic field strengths. Placing the theory on a manifold with boundary induces an action on 3d theories with $U(1)^n$ global symmetry. We first review the bulk action and then derive the corresponding boundary operations, following Refs.~\cite{DiPietro:2019hqe,Dimofte:2011ju}.

First we discuss how $\Sp(2n,\bZ)$ acts on the 4d theory. The action of $n$ Abelian $\U(1)$ gauge fields $A^I$ where $F^I = dA^I$ with complexified gauge couplings $\tau_{IJ}$ is
\begin{align}
    S_M[A^I,\tau_{IJ}] = \frac{1}{4\pi}\int_{X} \tau_{IJ}F^{-I} F^{-J} + \bar{\tau}_{IJ}F^{+I} F ^{+J}    \, ,
\end{align}
where $\tau_{IJ} = \frac{\theta_{IJ}}{2\pi}+\frac{2\pi \i}{g_{IJ}^2}$ and $F^{\pm I}=\frac12(F^I \mp \i \star F^I)$. $\Sp(2n,\bZ)$ acts on the couplings as
\begin{align}
    \tau_{IJ}' = (A_I^L \tau_{LM} + B_{IM})(C^{JN}\tau_{NM} + D^J_M)^{-1} \, ,
\end{align}
where
\begin{align}
    M = \begin{pmatrix}
    A & B\\
    C & D
    \end{pmatrix} \in \Sp(2n,\bZ) \, .
\end{align}

There are 3 types of matrices that the generators fall into \cite{hua1949generators}. The familiar ones are the T type and S type 
\begin{align}
   \text{T-type}:& \begin{pmatrix}
    I & B\\
    0 & I
    \end{pmatrix} \, , \, \text{B is the symmetric matrix generating $\tau \to \tau + B$} \\
    \text{S-type}:& \begin{pmatrix}
    I - J & J\\
    -J & I - J 
    \end{pmatrix} \, , \, \text{$J=\text{diag}(j_1,j_2,\cdots,j_n)$ and $j_i \in \{0,1\}$} \, .
    \label{eq:S-J}
\end{align}
For the S-type transformations, $j_i=1$'s gauge the fields $A_i$. The additional ingredient that we did not have for $n=1$ is a type of transformation that rotates the fields into each other. These ``GL-Type" transformations act in the following way:
\begin{align}
   \text{GL-type}:& \begin{pmatrix}
    U & 0\\
    0 & (U^{-1})^T
    \end{pmatrix} \, , \text{where $U \in \mathrm{GL}(n,\mathbb Z)$ generates $A'=U^T A$} \, .
\end{align}

\subsection{Induced operations on the 3d boundary}

Let us couple the 4d theory $S[A^I,\tau_{IJ}]$ to a theory living on the boundary such that the total action is
\begin{align}
    S_M[A^I,\tau_{IJ}] + \frac{1}{2\pi}\int_{\partial X}\star j_I A^I\, ,
    \label{eq:full-action}
\end{align}
where $j_I$ is the 3d current. We explicitly take $X$ to be half space in Minkowski spacetime with boundary $\partial X$.

Acting with a T-type transformation we find 
\begin{align}
    S_M[A^I,\tau_{IJ}+B_{IJ}] &+ \frac{1}{2\pi}\int_{\partial X}\star j_I  A^I = S_M[A^I,\tau_{IJ}] + \frac{1}{2\pi}\int_{\partial X}\star j_I  A^I +\frac{1}{4\pi}\int_X B_{IJ}F^I F^J \\
    &=S_M[A^I,\tau_{IJ}] + \frac{1}{2\pi}\int_{\partial X}\star j_I  A^I +\frac{1}{4\pi} B_{IJ}A^I dA^J \, ,
\end{align}
so the T-type transformations act by stacking a Chern-Simons or BF theory in 3d.\footnote{Our derivation slightly differs from \cite{Seiberg:2016gmd} where they wanted the action to be a genuine symmetry of the bulk so they defined the T-type transformation to compensate with a level $k=-1$ Chern-Simons term on the boundary. The S-type transformation also contributed $-\frac{1}{2\pi}adA$ to the boundary, which was addressed by redefining $S\mapsto -S$. One could make a similar argument for the GL-type transformation but we choose to just inherit the boundary rotation of $A$'s from the bulk theory.}

For the S type transformations, we can almost exactly follow \cite{Seiberg:2016gmd}. First add 
\begin{align}
    +\frac{1}{2\pi}\int_X J_
    {IJ}F'^{I} (F - dA)^J
\end{align}
to the action Eq. \eqref{eq:full-action} where $F'^I$ and $F$ are arbitrary two forms and $J$ is defined in Eq. \eqref{eq:S-J}. As a path integral over $F'^I$ just enforces $F^I=dA^I$, we can treat $F^I$ independent in the original action and integrate over $F^I$ first while keeping the boundary value fixed. After doing the Gaussian integral, we are left with 
\begin{align}
    S_M[A'^I,\tau'_{IJ}] + \frac{1}{2\pi}\int_X J_{IJ}F'^Id A^J+ \frac{1}{2\pi}\int_{\partial X}\star J_I A^I \, ,
\end{align}
where 
\begin{align}
    \tau' = ((I-J)\tau + J)(-J\tau + (I-J))^{-1} \, .
\end{align}
Note the last term still explicitly depends on $A^I$ and we cannot express it in terms of an arbitrary $F^I$. Locally we find $F'^I = d A'^I$.

Denoting the boundary value of $A^I$ as $a^I$ and applying Stokes theorem, we have 
\begin{align}
    S_M[A'^I,\tau'_{IJ}] + \frac{1}{2\pi}\left(\int_{\partial X} \star J_I a^I + J_{IJ}A'^Ida^J\right) \, .
\end{align} 

Thus we have found that in 3d, the $S$-type transformations gauge the $\U(1)$ symmetry for fields $A^I$ with $j_I=1$.

To derive the GL-type transformations, note that for $\tau' = U\tau U^T$,
\begin{align}
    S_M[A,U\tau U^T] = S_M[U^T A,\tau] \, ,
\end{align}
and we will have the boundary theory inherit the same rotation from the bulk.

All together, these operations are realized on our 3d theory $Z[A]$ as

\begin{align}
   \text{T-type}:&~ Z[A]\to \exp{\left(\frac{\i}{4\pi}\int B_{IJ}A^I dA^J\right)}Z[A] \\
   \text{S-type}:&~ Z[A]\to \int \cD a~\exp{\left(\frac{\i}{2\pi}\int J_{IJ}a^I dA^J\right)}Z[Ja+(I-J)A] \\
   \text{GL-type}:&~ Z[A]\to Z[U^TA] \, .
\end{align}

\subsection{$\Sp(4,\bZ)$ operations for $U(1)^2$}

We now specialize to $n=2$. The complex symmetric coupling matrix belongs to the genus-two Siegel upper half-space $\mathbb H_2$, and the quantum parameter space is $\mathbb H_2/\Sp(4,\bZ)$. Following Ref.~\cite{DiPietro:2019hqe}, we use the following generators\footnote{Note that $\sS,\sT,\sR_1$ and $\sR_2$ generate $\Sp(4,\bZ)/\sim$ where $\sS \sim -\sS$ of $\Sp(4,\bZ)$ and our definition of $\mathsf S$ differs from \cite{DiPietro:2019hqe} by a sign. Its central element $-I$ acts trivially on the coupling matrix $\tau$ but induces simultaneous charge conjugation of the boundary $U(1)^2$ backgrounds.} :
\begin{equation}
    \sS = \begin{pmatrix}
    0 & 0 & 1 & 0\\
    0 & 1 & 0 & 0\\
    -1 & 0 & 0 & 0\\
    0 & 0 & 0 & 1
    \end{pmatrix}\, ,
    \label{eq:S-gen}
\end{equation}
\begin{equation}
    \sT = \begin{pmatrix}
    1 & 0 & 1 & 0\\
    0 & 1 & 0 & 0\\
    0 & 0 & 1 & 0\\
    0 & 0 & 0 & 1
    \end{pmatrix}\, ,
    \label{eq:T-gen}
\end{equation}
\begin{equation}
    \sR_1 = \begin{pmatrix}
    0 & 1 & 0 & 0\\
    1 & 0 & 0 & 0\\
    0 & 0 & 0 & 1\\
    0 & 0 & 1 & 0
    \end{pmatrix}\, ,
    \label{eq:R1-gen}
\end{equation}
\begin{equation}
    \sR_2 = \begin{pmatrix}
    1 & 1 & 0 & 0\\
    0 & 1 & 0 & 0\\
    0 & 0 & 1 & 0\\
    0 & 0 & -1 & 1
    \end{pmatrix}\, .
    \label{eq:R2-gen}
\end{equation}
We now pass to the description where the duality group acts on the Lagrangian. It is worth keeping in mind that these operations still take place within the path integral where lowercase letters are dynamical fields and uppercase letters are background fields. The action of these generators on a 3d theory with $U(1)\times U(1)$ symmetry was first found in \cite{Dimofte:2011ju}.  They are:
\begin{equation}
    S: \mathcal{L}(A_1,A_2) \mapsto \mathcal{L}(a_1,A_2) + \frac{1}{2\pi}a_1 dA_1 \, ,
\end{equation}
\begin{equation}
    T: \mathcal{L}(A_1,A_2) \mapsto \mathcal{L}(A_1,A_2) + \frac{1}{4\pi}A_1 dA_1\, ,
\end{equation}
\begin{equation}
    R_1: \mathcal{L}(A_1,A_2) \mapsto \mathcal{L}(A_2,A_1)\, ,
\end{equation}
\begin{equation}
    R_2: \mathcal{L}(A_1,A_2) \mapsto \mathcal{L}(A_1,A_1+A_2) \, .
\end{equation}
It is convenient to consider $S$ and $T$ transformations simultaneously on both copies. For the first $S$ transformation, we want it to act as 
\begin{equation}
    S_d : \mathcal{L}(A_1,A_2) \mapsto \mathcal{L}(a_1,a_2) + \frac{1}{2\pi}a_1 dA_1 + \frac{1}{2\pi}a_2 dA_2 \, ,
    \label{eq:S_d}
\end{equation}
a diagonal gauging which can be obtained from the generators by $S_d:=SR_1SR_1$. 
For the second $S$ transformation, we desire
\begin{equation}
    S_o: \mathcal{L}(A_1,A_2) \mapsto \mathcal{L}(a_1,a_2) + \frac{1}{2\pi}a_1 dA_2 + \frac{1}{2\pi}a_2 dA_1 \, ,
    \label{eq:S_o}
\end{equation}
an off diagonal gauging which can be defined as $S_o:=R_1S_d$.
This is accomplished by
\begin{equation}
    \sS_d = \left(
\begin{array}{cccc}
 0 & 0 & 1 & 0 \\
 0 & 0 & 0 & 1 \\
 -1 & 0 & 0 & 0 \\
 0 & -1 & 0 & 0 \\
\end{array}
\right) ;\,\sS_o = \left(
\begin{array}{cccc}
 0 & 0 & 0 & 1 \\
 0 & 0 & 1 & 0 \\
 0 & -1 & 0 & 0 \\
 -1 & 0 & 0 & 0 \\
\end{array} \right) \,.
\label{eq:Sd-matrix}
\end{equation}

Analogous to $S_d$ and $S_o$, we can define $T_d:=TR_1TR_1$ and $T_o:=TT_dR_2^{-1}T_d^{-1}R_2$\footnote{On the boundary, inverse actions of the T-type transformations correspond to stacking the relevant Chern-Simons terms or BF terms with the wrong sign. The inverse action of the rotation $R_2$ is $R_2^{-1}: \cL(A_1,A_2) \mapsto \cL(A_1,A_2-A_1)$. Also see Appendix \ref{app:presentation}. }which acts as
\begin{align}
    T_d: \mathcal{L}(A_1,A_2) \mapsto \mathcal{L}(A_1,A_2) + \frac{1}{4\pi}A_1 dA_1 + \frac{1}{4\pi}A_2 dA_2 \, ,
\end{align}
and 
\begin{align}
    T_o: \mathcal{L}(A_1,A_2) \mapsto \mathcal{L}(A_1,A_2) + \frac{1}{2\pi}A_1 dA_2\, .
\end{align}
The matrix expressions can be written as 
\begin{equation}
    \sT_d = \left(
\begin{array}{cccc}
 1 & 0 & 1 & 0 \\
 0 & 1 & 0 & 1 \\
 0 & 0 & 1 & 0 \\
 0 & 0 & 0 & 1 \\
\end{array}
\right) ;\,\sT_o = \left(
\begin{array}{cccc}
 1 & 0 & 0 & 1 \\
 0 & 1 & 1 & 0 \\
 0 & 0 & 1 & 0 \\
 0 & 0 & 0 & 1 \\
\end{array} \right) \,.
\label{eq:Td-matrix}
\end{equation}

For the $n=1$ case, we had $S^2=-1$ (or charge conjugation) and $(\sS \sT)^3=1$ in the bulk, but on the boundary $(ST)^3: \cL[A]\rightarrow \cL[A]+ U(1)_1$, i.e. a level one Chern-Simons theory attached to the original theory. A natural question to ask is how this carries over to $S_d^2, S_o^2$, $(S_o T_o)^3$, and $(S_d T_d)^3$. At the level of the matrix products using Eq. \eqref{eq:Sd-matrix} and Eq. \eqref{eq:Td-matrix}, we obtain $\sS_d^2=\sS_o^2=-1$ and $(\sS_o \sT_o)^3=(\sS_d \sT_d)^3=1$. Below we check these expressions at the level of the 3d Lagrangian.

For $S_d^2$, we can apply $S_d$ once more to Eq. \eqref{eq:S_d} to get
\begin{align}
    S_d:\cL'(A_1,A_2) \mapsto \cL(a_1,a_2) + \frac{1}{2\pi}a_1 db_1 + \frac{1}{2\pi}a_2 db_2 + \frac{1}{2\pi}b_1 dA_1 + \frac{1}{2\pi}b_2 dA_2 \, ,
\end{align}
where $\cL'(A_1,A_2)$ is the RHS of Eq. \eqref{eq:S_d}. As a result, we can simplify this expression to 
\begin{align}
S_d^2:\cL(A_1,A_2)\mapsto&\cL(a_1,a_2) + \frac{1}{2\pi}b_1 d(a_1 + A_1) + \frac{1}{2\pi}b_2 d(a_2 + A_2) \\
    &=\cL(-A_1,-A_2) \, ,
\end{align}
where we integrated by parts in the first line and the path integral over $b_i$ just gives $\delta(a_i+A_i)$. Then the integral over $a_i$ sets $a_i = -A_i$ giving us $S_d^2:\cL(A_1,A_2) \mapsto \cL(-A_1,-A_2)$. Similarly, we can show applying $S_o$ to the RHS of Eq. \eqref{eq:S_o} gives
\begin{align}
    S_o:\cL'(A_1,A_2) \mapsto&\cL(a_1,a_2) + \frac{1}{2\pi}b_1 d(a_2 + A_2) + \frac{1}{2\pi}b_2 d(a_1 + A_1) \\
    &=\cL(-A_1,-A_2) \, ,
\end{align}
so we see $S_o^2:\cL(A_1,A_2) \mapsto \cL(-A_1,-A_2)$. It is easy to see $S_o^4=S_d^4=1$. The above is not surprising and is consistent with matrix multiplication.

Of more interest are the $(S_dT_d)^3$ and $(S_oT_o)^3$ actions. For $(S_dT_d)^3$, we have 
\begin{align}
    (S_dT_d)^3:&\cL[A_1,A_2] \mapsto \cL[a_1,a_2]+\sum_{i=1}^2\left(\frac{1}{4\pi}a_i da_i + \frac{1}{2\pi}a_i db_i + \frac{1}{4\pi} b_i db_i + \frac{1}{2\pi}b_i dc_i + \frac{1}{4\pi}c_idc_i + \frac{1}{2\pi}c_i dA_i\right) \\
    &=\cL[a_1,a_2]+\sum_{i=1}^2\left(\frac{1}{4\pi}(a_i+b_i+c_i)d(a_i+b_i+c_i)-\frac{1}{2\pi}a_idc_i + \frac{1}{2\pi}c_i dA_i\right) \, .
\end{align}
Let us define $\tilde a_i:=a_i+b_i+c_i$ and do the integral over $c_i$ which sets $a_i=A_i$ so,
\begin{align}\label{eq:sdtd3}
    (S_dT_d)^3:\cL[A_1,A_2] &\mapsto \cL[A_1,A_2]+ \frac{1}{4\pi}\tilde a_1 d \tilde a_1 + \frac{1}{4\pi}\tilde a_2 d \tilde a_2 \, .
\end{align}

Thus we have identified a projective phase for $\Sp(4,\bZ)$ analogous to the one in $\SL(2,\bZ)$, just two dynamical level one Chern-Simons theories. 

Similarly, we can look at $(S_oT_o)^3$ where 
\begin{align}
    (S_o T_o)^3:\cL[A_1,A_2] &\mapsto \cL[a_1,a_2]+ \frac{1}{2\pi}a_2 da_1 + \frac{1}{2\pi}a_2 db_1 + \frac{1}{2\pi}a_1db_2 +\frac{1}{2\pi}b_1db_2 \\&+ \frac{1}{2\pi}b_1dc_2 + \frac{1}{2\pi}b_2dc_1 
    + \frac{1}{2\pi}c_1 dc_2 + \frac{1}{2\pi}c_1 d A_2 + \frac{1}{2\pi}c_2 d A_1 \, ,
\end{align} 
and after rewriting things in terms of $\tilde a_i$ and integrating out $c_i$, we obtain 
\begin{align}
    (S_o T_o)^3:\cL[A_1,A_2] &\mapsto \cL[A_1,A_2]+ \frac{1}{2\pi}\tilde a_1 d \tilde a_2 \, .
\end{align} 
where $\frac{1}{2\pi}\tilde a_1 d \tilde a_2$ corresponds to a trivial theory \cite{Witten:2003ya}, and it cannot be identified as a projective action.

\subsection{Projective action and an order-five duality of $U(1)^2$}
As mentioned before, for the $U(1)$ case, the $\SL(2,\bZ)$ symmetry acts projectively on the $3d$ boundary, following \eqref{eq:proj-rep},
\begin{equation}
    (ST)^3 : \cL[A] \rightarrow \cL[A]+\frac{1}{4\pi}ada
\end{equation}
where $\frac{1}{4\pi}ada$ corresponds to the decoupled $U(1)_1$ theory with chiral central charge $1$. This projective action is related to the $\bZ_{12}$ anomaly of the $\SL(2,\bZ)$ duality symmetry of the $4d$ Maxwell theory. Details are presented in App.~\ref{app:u1anomaly}.

For $U(1)^2$, naive projective actions can be built from the single $\SL(2,\bZ)$ transformation, e.g. \eqref{eq:sdtd3}. Interestingly, there is an order-5 duality transformation that intrinsically mixes the two components and has projective action on the boundary for $U(1)^2$, which goes beyond the simple construction. We define,
\begin{equation}
    g_5=T R_2 R_1 S:\;\cL[A_1,A_2]\rightarrow \cL[a,A_1]+\frac{1}{4\pi}A_1dA_1+\frac{1}{2\pi}ad(A_1+A_2)
\end{equation}
and as $4\times 4$ matrix, $\sg_5^5 = I$ which generates the $\bZ_5$ subgroup of $\Sp(4,\bZ)$ \footnote{The largest prime order subgroup of $\Sp(4,\bZ)$ is this $\bZ_5$. More generally for $\Sp(2n,\bZ)$, the prime order $p$ is the largest prime that $p-1\le2n$ \cite{burgisser1982elements}.}, and,
\begin{equation}
    g_5^5:\; \cL[A_1,A_2]\rightarrow \cL[A_1,A_2]+\frac{1}{4\pi}ada
\end{equation}
where again $\frac{1}{4\pi}ada$ corresponds to the decoupled $U(1)_1$ theory. The derivation is straightforward but tedious, we leave it to the App.~\ref{app:u1anomaly}. As for $U(1)$, this projective action suggests an anomaly of the duality symmetry. Since $\sg_5^5 = 1$ and acts on boundary as $g_5^5=X^2$, where $X$ is the operator to stack a layer of fermionic invertible phase $p+ip$ superconductor. By redefining $g_5'=X^mg_5 $, $(g'_5)^5=X^{5m+2}$, the projective phase cannot be removed. In general, the projectivity for some duality group $D$ on a spin three manifold is classified by    $H^2(BD,\operatorname{Inv}_{\text{spin}}^3)$, where $\operatorname{Inv}_{\text{spin}}^3\cong\bZ$ is the group of invertible phases \cite{Bhardwaj:2020ymp}. So for $D=\bZ_5$, it gives the $2\in \bZ_5 \cong H^2(\bZ_5,\bZ)$. This is the direct order-five analogue of the single-$U(1)$ relation $(\widehat S\widehat T)^3=X^2$.
Moreover, the anomaly should be seen by the image of $g_5$ under the abelianization of the duality group. As pointed out in \cite{Seiberg:2018ntt}, although the Maxwell duality transformations factor through $\Sp(4,\bZ)$, the projective anomaly can be studied more sharply after pulling the action back to the genus-two mapping class group $\Gamma_2$, for which $H^2(B\Gamma_2,\bZ)\cong\bZ_{10}$, while $\Sp(4,\bZ)$ has the anomaly $\bZ\oplus \bZ_2$ \cite{benson2018cohomology}. Following this logic, the anomaly index $8\in \bZ_{10}$ restricts to the anomaly $2\in \bZ_5$ of the $\bZ_5$ subgroup matches with our calculation, suggesting the $U(1)^2$ has anomaly index $8\in \bZ_{10}$ in comparison to the single Maxwell theory with $8\in \bZ_{12}\cong H^2(\SL(2,\bZ),\bZ)$ \cite{Seiberg:2018ntt}. The details are presented in App.~\ref{app:u1anomaly}.

\section{$K$-matrices and symmetric theories}
\label{sec:kmatrix}
In this section, we specialize to the case of field theories with $U(1)$ $\times$ $U(1)$ global symmetry and, in particular, discuss the theories that we can generate with the $\Sp(4,\mathbb{Z})$ transformations described above. 
In 2+1d, given the anomaly-free continuous symmetry $G$, besides the spontaneously symmetry breaking phase with Goldstone modes, its symmetric gapped phases can be understood as the $G$-enriched topological phases \cite{barkeshliSymmetryFractionalizationDefects2019a}, with $G$ symmetric trivial phase being the simplest case. When the topological part is given by Abelian topological order, we can use the $K$-matrix to describe them \cite{Wang:2020nmz}. 
We begin by reviewing how $K$-matrices, which we use extensively in section four, describe the low energy physics of Abelian topological phases of matter.  We will discuss how to couple such phases to an extrinsic zero-form symmetry, which will enable us to use the technology in the previous section to generate a web of theories describing a variety of symmetric gapped\footnote{Up to occasional Nambu-Goldstone modes.} phases.  We then warm up with an account of the theories generated by $\SL(2,\mathbb{Z})$ transformations before turning to those generated by $\Sp(4,\mathbb{Z})$ transformations.
\subsection{Review of $K$-matrix formalism}
The Abelian Chern--Simons theory is specified by a nondegenerate integral symmetric matrix \(K\) \footnote{Depending on the $K$-matrix, the topological order of such a theory might be trivial.},
\begin{equation}
    \mathcal L=\frac{1}{4\pi}K_{IJ}a^I da^J .
\end{equation}
There are two cases that should be distinguished. If \(K\) is even, meaning
\begin{equation}
    \Lambda^T K\Lambda\in 2\mathbb Z
    \qquad
    \text{for every }\Lambda\in\mathbb Z^N,
\end{equation}
or equivalently if all diagonal entries \(K_{II}\) are even, the action defines a bosonic Chern--Simons theory on an oriented three-manifold. If \(K\) is odd, the theory depends on a choice of spin structure and defines a spin Chern--Simons theory. Wilson lines are labeled by vectors \(\ell\in\mathbb Z^N\). After
identifying lines that differ by a local excitation,
\(\ell\sim\ell+K\Lambda\), the fractional quasiparticle sectors are
labeled by
\begin{equation}
    \mathcal A_K=\mathbb Z^N/K\mathbb Z^N,
    \qquad
    |\mathcal A_K|=|\det K|.
\end{equation}
The mutual braiding phase
\begin{equation}
    M_{\ell,\ell'}
    =\exp\!\left(2\pi i\,\ell^T K^{-1}\ell'\right)
\end{equation}
is well defined on \(\mathcal A_K\) for any integral \(K\).

For even \(K\), the topological spin is also well defined on
\(\mathcal A_K\):
\begin{equation}
    h_{\ell}
    =\frac{1}{2}\ell^T K^{-1}\ell
    \pmod 1,
    \qquad
    \theta_{\ell}
    =\exp\!\left(2\pi i h_{\ell}\right)
    =\exp\!\left(\pi i\,\ell^T K^{-1}\ell\right).
\end{equation}
Since the theory is Abelian, its total quantum dimension is
\begin{equation}
    \mathcal D=\sqrt{|\det K|}.
\end{equation}
The chiral central charge of this \(K\)-matrix realization is
\begin{equation}
    c_-=\operatorname{sig}(K),
\end{equation}
where \(\operatorname{sig}(K)=n_+-n_-\), the difference between positive and negative eigenvalues. For even \(K\), this is consistent with the ordinary Gauss-Milgram formula
\begin{equation}
    \frac{1}{\sqrt{|\det K|}}
    \sum_{\ell\in\mathcal A_K}
    \exp\!\left(\pi i\,\ell^T K^{-1}\ell\right)
    =
    \exp\!\left(\frac{2\pi i}{8}\operatorname{sig}(K)\right).
\end{equation}

For odd \(K\), the expression
\(\frac12\ell^T K^{-1}\ell\) does not define a function
\(\mathcal A_K\to\mathbb Q/\mathbb Z\). Indeed,
\begin{equation}
\begin{split}
    h_{\ell+K\Lambda}-h_\ell
    &=
    \ell^T\Lambda
    +\frac12\Lambda^T K\Lambda ,
\end{split}
\end{equation}
which can be half-integral when \(K\) is odd. Thus, after quotienting
by local fermions, \(h_\ell\) is defined only modulo \(1/2\), and \(\theta_\ell\) is defined only up to a sign. The ordinary Gauss-Milgram sum written above therefore must not be applied directly to \(\mathcal A_K\). Nevertheless, the \(K\)-matrix realization still has $ c_-=\operatorname{sig}(K)$.

An anyon theory $\cC$ \footnote{By anyon theory, we mean the modular tensor category \cite{Kitaev:2005hzj} that contains information about the anyons, their fusion, and their braiding.} can be enriched by the $0$-form global symmetry $G$, yielding a symmetry enriched topological order (SET) \cite{barkeshliSymmetryFractionalizationDefects2019a}, whose topological data is given by the $G$-crossed braided tensor category.

When the obstruction classes $H^4(G,U(1))$ and $H^3_{\rho}(G,\mathcal{A})$ vanish, the $G$-crossed braided tensor category contains three layers of topological data. The first is the symmetry action $[\rho]: G \to \mathrm{Aut}(\cC)$, which specifies how the global symmetry permutes the anyon types of the underlying modular tensor category $\cC$. The second is the symmetry fractionalization class $[\mathfrak{w}] \in H^2_\rho(G, \cA)$, valued in the group of Abelian anyons $\cA$ with coefficients twisted by $\rho$, which characterizes how anyons carry fractional quantum numbers of $G$. The third is a class $[\alpha] \in H^3(G, \mathrm{U}(1))$, which labels distinct ways of assigning consistent $F$-symbols to the symmetry defects; physically it corresponds to the freedom of stacking a $G$-SPT phase onto the topological order without changing either the symmetry action or the fractionalization pattern.

For the bilayer quantum Hall states considered later, the relevant symmetry group is $G = \mathrm{U}(1)_c \times \mathrm{U}(1)_s$ (or a subgroup thereof together with the discrete layer-exchange symmetry $\bZ_2^X$), and much of the hierarchy structure can be understood in terms of the fractionalization data.

We now discuss the simplest case, where we enrich the $K$-matrix theory by $U(1)$ zero-form symmetry. The fractionalization data together with possible background Chern-Simons term specifies the $U(1)$ enrichment as $U(1)$ cannot permute anyons. We couple the $K$-matrix theory to the background $U(1)$ gauge field $A$,
\begin{equation}
    \mathcal{L}_{coupling} = \frac{1}{2\pi}t_I a^I dA,
\end{equation}
and the $t^I$ specifies the symmetry fractionalization class \cite{Lu:2013jqa,Hung:2013nla,Cheng:2022nds}. Physically, threading a $2\pi$ flux of the external gauge field $A$ will pump an Abelian anyon $e^{i \oint t^I a^I}$. The anyons then carry the $U(1)$ charge $t^\intercal K^{-1} l_q \mod 1$, where $l_q$ labels the anyon $e^{i \oint l_q^I a^I}$. When integrating out the dynamical gauge field $a^I$, we get the effective Lagrangian,
\begin{equation}
    -\frac{(K^{-1})_{IJ}t_I t_J}{4\pi}AdA
\end{equation}
and we define the Hall conductivity to be \footnote{This is to match the ordinary convention, one can also change the convention of $K$-matrix with an additional minus sign.}
\begin{equation}
\sigma_{xy} = (K^{-1})_{IJ}t_I t_J.  
\end{equation}
The fractional charge of the anyon $\ell$ under the symmetry $U(1)$ is
\begin{equation}
    Q(\ell)
    =
    t^{T}K^{-1}\ell
    \quad \mathrm{mod}\; \mathbb Z .
\end{equation}
The freedom to shift $Q(\ell)$ by an integer follows from the quotient used to define $\mathcal{A}$.  In the following two subsections, we examine the $K$-matrices that result from applying the transformations studied above to $A$.

\subsection{$U(1)$ and $U(1)\times U(1)$ symmetric theories}
We now apply the manipulation in the previous section to the $U(1)$ background $A$. The action of $S$ and $T$ transformation generates additional $U(1)$ gauge fields and Chern-Simons terms associated to them. The Lagrangian is generally of the form
\begin{equation}
    \mathcal{L} = a^1 \star j +\frac{K_{IJ}}{4\pi}a^I d a^J+\frac{m}{4\pi}AdA+\frac{t}{2\pi}a^N d A.
\end{equation}

In the App.~\ref{app:mat}, we discuss how to encapsulate this Lagrangian as an enlarged $K$-matrix, for now, we turn to the phases that can be encapsulated by Lagrangians of this sort.  In the previous subsection, we described how the data of SET phases are encoded in $K$ and $t$.  We will see many examples of such phases in the following section, so let us consider two more prosaic phases.  The first, which results from applying $T$ to the trivial theory is an invertible phase with Lagrangian
\begin{equation}
    \mathcal{L}_{IQH} = \frac{k}{4\pi} A dA,
\end{equation}
which describes an integer quantum Hall phase with Hall conductivity $k$ in appropriate units.  The second, which results from applying $S$ to the trivial theory, has the Lagrangian
\begin{equation}
    \mathcal{L}_{SSB} = \frac{1}{2\pi}a dA.
\end{equation}
This describes a phase that spontaneously breaks the entire $U(1)$ symmetry.  The Nambu-Goldstone boson is simply the dual photon \footnote{Note that if we included the less relevant Maxwell term for $a$, we would recover the discussion in i.e. \cite{Turner:2019wnh}.}.  Since there are infinitely many gapped phases, we make no attempt to work out an exhaustive web as in i.e. \cite{Moradi:2022lqp,Huang:2023pyk,Karch:2025hsj,Spieler:2025hyr}. 

We now turn to $U(1) \times U(1)$ symmetric theories.  The general Lagrangian is now 
\begin{align}
    \mathcal{L} = &\star j a^1 + \frac{K_{IJ}}{4\pi} a^I da^J + \frac{m_1}{4\pi}A_1 dA_1 + \frac{m_2}{4\pi}A_2 dA_2 \nonumber\\ &+\frac{n}{2\pi}A_1dA_2+ \frac{t_{1I}}{2\pi}a^I dA_1 + \frac{t_{2I}}{2\pi}a^I dA_2.
\end{align}

Special cases of the above Lagrangian encode a variety of symmetric phases. As before, we leave the discussion of the topologically ordered phases to the later section. Of course, we can have an integer quantum Hall phase characterized by response to the first or second copy of $U(1)$:  
\begin{equation}
    \mathcal{L}_{IQH1} = \frac{k_1}{4\pi}A_1dA_1,
\end{equation}
\begin{equation}
    \mathcal{L}_{IQH2} = \frac{k_2}{4\pi}A_2 dA_2.
\end{equation}
We can also break the first or second copy of $U(1)$:
\begin{equation}
    \mathcal{L}_{SSB1} = \frac{1}{2\pi} a_1 dA_1,
\end{equation}
\begin{equation}
    \mathcal{L}_{SSB2} = \frac{1}{2\pi}a_2 dA_2,
\end{equation}
or we can break the entire $U(1)\times U(1)$ symmetry:
\begin{equation}
    \mathcal{L}_{SSB} = \frac{1}{2\pi} a_1 dA_1 + \frac{1}{2\pi}a_2 dA_2.
\end{equation}
Having two copies of $U(1)$ lets us pair the two backgrounds.  This yields an SPT phase:
\begin{equation}
    \mathcal{L}_{SPT} = \frac{n}{2\pi} A_1 dA_2.
\end{equation}
There are also SSB phases that break more interesting subgroups of $U(1) \times U(1)$.  We have
\begin{equation}
    \mathcal{L}_{SSBD} = \frac{1}{2\pi} a_1 (dA_2+dA_1),
\end{equation}
which breaks the diagonal subgroup of $U(1)\times U(1)$, and
\begin{equation}
    \mathcal{L}_{SSBD'} = \frac{1}{2\pi} a_1 (dA_2 -dA_1),
\end{equation}
which breaks the off-diagonal subgroup.
\section{Application to bilayer quantum Hall}
\label{sec:bilayer}
The modern hierarchy picture of the fractional quantum Hall effect grew out of the observation that the quasiparticles of a Laughlin fluid can themselves form new incompressible fluids. Haldane and Halperin formulated this as an iterative construction of daughter states, naturally giving continued-fraction filling fractions \cite{Haldane:1983xm,Halperin:1984zz}. In parallel, Halperin's multicomponent wave functions introduced the \((m_1,m_2,n)\) states that later became the standard language for spinful and bilayer quantum Hall system \cite{Halperin:1983mmn}.

The Abelian Chern--Simons and \(K\)-matrix formulation, developed by Blok--Wen, Read, Wen--Zee and Wen, uses \(K\)-matrices to describe the hierarchy states \cite{Blok:1990mc,Read:1990zza,Wen:1992uk,Wen:1995qn}; see also the review \cite{Tong:2016kpv}. Jain's composite-fermion construction gives a complementary microscopic construction: attaching \(2p\) vortices to each electron maps the prominent fractions to integer quantum Hall states of composite fermions, producing the sequences \(\nu=n/(2pn\pm1)\) \cite{Jain:1989zz}. For the Abelian Jain sequences, these two descriptions should not be viewed as competing pictures. The composite-fermion language gives economical trial wave functions and an effective-field intuition, while the hierarchy language records the same topological order through a \(K\)-matrix, often after a change of basis. 

For bilayer quantum Hall systems the same relation becomes richer because there are separate layer charges, a possible layer-exchange symmetry, and both intralayer and interlayer correlations. The Halperin states are the natural bilayer parents, while bilayer composite-fermion constructions allow flux attachment both within and between layers \cite{scarola2001phase,Nguyen:2024jip}. Consequently a filling fraction alone does not uniquely identify the phase, while the additional pseudospin vector and its fractionalization data are part of the physical distinction.

In the following, we first revisit the single component hierarchy state using $\SL(2,\bZ)$ transformation, then generalize it to bilayer hierarchy state using the \(\Sp(4,\bZ)\) action.

\subsection{Hierarchies and $\SL(2,\bZ)$}
In the \(K\)-matrix formalism, the Haldane--Halperin hierarchy \cite{Haldane:1983xm,Halperin:1984zz} has a natural Abelian Chern--Simons description \cite{Blok:1990mc,Read:1990zza,Wen:1992uk,Wen:1995qn}. For a $N\times N$ $K$-matrix, the anyon group is \(\cA_K=\bZ^N/K\bZ^N\), with an anyon represented by an integer vector \(\ell\). Its fractional electric charge and mutual braiding with another anyon \(\ell'\) are
\begin{equation}
    q_\ell=t^\intercal K^{-1}\ell \mod 1\,,\qquad
    M_{\ell,\ell'}=\exp(2\pi i\,\ell^\intercal K^{-1}\ell') \,.
\end{equation}
The \(\U(1)\) symmetry fractionalization class is equivalently encoded by the distinguished anyon,
\begin{equation}
    v=[t]\in\cA_K\,,
\end{equation}
which is produced by inserting \(2\pi\) flux of the background electromagnetic \(\U(1)\) and is called \textit{vison}.  The $U(1)$ charge of the anyons is defined by its Aharonov-Bohm, or equivalently the braiding with the vison $v$ generated by inserting the \(2\pi\)-flux,
\begin{equation}
    M_{\ell,v}=\exp(2\pi i q_\ell)\,.
\end{equation}
In particular, the Hall conductivity \(\sigma_H\) is given by the charge of the vison,
\begin{equation}
    q_v=t^\intercal K^{-1}t=\sigma_H \,,
\end{equation}
where \(\sigma_H\) is in units of \(e^2/h\), modulo an integer. This is Laughlin's flux-insertion argument in topological language: threading \(2\pi\) flux through an annulus pumps charge \(\sigma_H\), which is simply the charge of the vison created by the flux. This identification is purely topological. The further statement that \(\sigma_H\) equals the filling fraction \(\nu\) follows from the Galilean invariance: in a uniform electric field the center of mass drifts with \(v_D^i=\epsilon^{ij}E_j/B_{\rm ext}\), independently of interactions, so for mean charge density \(\bar\rho\) the current is \(j^i=\bar\rho v_D^i=(\bar\rho/B_{\rm ext})\epsilon^{ij}E_j\). Since \(\nu=2\pi\bar\rho/B_{\rm ext}\), comparison with the Chern--Simons response \(j^i=(\sigma_H/2\pi)\epsilon^{ij}E_j\) gives \(\sigma_H=\nu\). This is the same center-of-mass kinematics underlying Kohn's theorem and the Galilean-invariant response theory of quantum Hall fluids \cite{Kohn:1961zz,Hoyos:2011ez}.

We now reinterpret the Haldane--Halperin hierarchy using the $\SL(2,\bZ)$ transformation. Early references have considered the relation between the $\SL(2,\bZ)$ transformation and composite fermion \cite{Burgess:2000kj}. Let us recall the elementary \(SL(2,\mathbb Z)\) operation that enters the hierarchy construction,
\begin{equation}
    \frac{1}{2\pi}adA
    \xrightarrow{\,ST^m\,}
    \frac{1}{2\pi}adb+\frac{m}{4\pi} bdb+\frac{1}{2\pi}bdA \,.
\end{equation}
Here \(ST^m=S\circ T^m\): one first adds the level-\(m\) Chern--Simons term \(\frac{m}{4\pi}AdA\), then gauges \(A\) by promoting it to a dynamical gauge field \(b\), and finally couples \(b\) to the new background field \(A\). Before the transformation,
\begin{equation}
    \star j_A=\frac{1}{2\pi}da \,,
\end{equation}
so a unit monopole operator \(\cM_a\) that threads \(2\pi\) flux of \(a\) carries \(\U(1)_A\) charge \(1\). After the transformation,
\begin{equation}
    \star j_A=\frac{1}{2\pi}db,\qquad da+m\,db+dA=0 \,.
\end{equation}
At vanishing background flux, a unit \(a\)-monopole therefore carries charge \(-1/m\). Thus this operation fractionalizes the unit charge to \(|Q_A|=1/m\), and this is the elementary step in the Abelian hierarchy construction.

Starting from a parent state \((K_0,t_0)\), tuning the magnetic field away from the parent filling creates quasiparticles or quasiholes, which may form a Laughlin-like daughter state. In field-theoretic terms, a hierarchy step can be organized as an \(ST^p\) operation on the background gauge field associated with the excess-flux current \cite{Witten:2003ya}. The \(T^p\) part adds a level-\(p\) Chern--Simons term, while the \(S\) part promotes the corresponding background field to a dynamical gauge field. In the Abelian Chern--Simons description this enlarges the \(K\)-matrix by one row and one column,
\[
    K_{n+1}
    =
    \begin{pmatrix}
        K_n & \lambda_n \\
        \lambda_n^\intercal & p
    \end{pmatrix},
    \qquad
    t_{n+1}
    =
    \begin{pmatrix}
        t_n \\ 0
    \end{pmatrix}\,.
\]
Here \(\lambda_n\in\bZ^n\) specifies the parent anyon that forms the daughter Laughlin state, and \(p\) gives the $1/p$ Laughlin state for the excessive quasiparticles/quasiholes. 

From the SET  point of view, each hierarchy step enlarges the $K$-matrix from \(\cA_{K_n}\) to \(\cA_{K_{n+1}}\), while the external electromagnetic \(\U(1)\) still couples through the original charge vector, \(t=(1,0,0,\ldots)\). The new dynamical gauge fields are not directly charged under the external \(\U(1)\), but they change the quotient \(\bZ^N/K\bZ^N\), and hence change the topological sector represented by the same vector \(t\). In this sense, the hierarchy construction refines the \(\U(1)\) symmetry fractionalization data. The daughter Hall conductivity is again the charge of the symmetry-flux anyon, \(\sigma_H^{\mathrm{daughter}}=t^\intercal K^{-1}t \mod 1\). A modern formulation of the hierarchy in terms of anyon condensation, including non-Abelian parent states, is given in \cite{Zhang:2024bye}.

To be specific, start from a bosonic or fermionic Laughlin state at filling \(\nu=1/m_1\),
\begin{equation}
    \frac{m_1}{4\pi} ada +\frac{1}{2\pi}adA \,,
\end{equation}
where \(A\) is the background gauge field for charge conservation. The gauge-invariant monopole operator \(\cM_a(\cM_A^\dagger)^{m_1}\) carries one unit of electric charge, so threading \(2\pi\) external flux pumps charge \(1/m_1\), as in Laughlin's argument.

At fixed density, \(\nu\propto 1/B_{\rm ext}\), so decreasing the magnetic field increases the filling and creates quasiparticles. We describe this branch by shifting
\begin{equation}
    dA\to dA-dB \,,
\end{equation}
and then letting the quasiparticles form a daughter Laughlin state by acting with \(ST^{m_2}\) on \(B\):
\begin{equation}
    \frac{m_1}{4\pi} ada +\frac{1}{2\pi}ad(A-B)
    \xrightarrow{ST^{m_2}\text{ on }B}
    \frac{m_1}{4\pi} ada +\frac{1}{2\pi}ad(A-b)
    +\frac{m_2}{4\pi}bdb+\frac{1}{2\pi}bdB \,.
\end{equation}
After setting the auxiliary background field \(B\) to zero, this gives the first hierarchy state
\begin{equation}
    K= \begin{pmatrix}
        m_1&-1\\
        -1 & m_2
    \end{pmatrix},\qquad t= (1,0),
\end{equation}
whose Hall conductivity is
\begin{equation}
    \sigma_H=t^\intercal K^{-1}t=\frac{m_2}{m_1m_2-1}\,.
\end{equation}

This procedure can be iterated recursively. Replacing \(B\to B-C\) and acting with \(ST^{m_3}\) on \(C\) gives
\begin{equation}
    \frac{m_1}{4\pi} ada +\frac{1}{2\pi}ad(A-b)
    +\frac{m_2}{4\pi}bdb+\frac{1}{2\pi}bd(B-c)
    +\frac{m_3}{4\pi} cdc +\frac{1}{2\pi} cdC \,.
\end{equation}
After the auxiliary background fields \(B,C,\cdots\) are set to zero, one obtains the tridiagonal hierarchy matrix
\begin{align}
    K= \begin{pmatrix}
    m_1 & -1 &  0& \dots& 0 \\
    -1 & m_2  & -1 & \\
    0 & -1   & m_3 & \ddots \\
    \vdots & &\ddots &\ddots& -1 \\
    0 & & & -1 & m_n
    \end{pmatrix},\qquad t=(1,0,\cdots,0) \,.
\end{align}
For fermionic hierarchy states, \(m_1\) is odd and \(m_i\) with \(i>1\) are even; for bosonic hierarchy states, \(m_1\) and all \(m_i\) with \(i>1\) are even. Choosing \(m_1=2p+1\) and \(m_i=2\) for \(i>1\) gives the fermionic Jain states
\begin{equation}
    \nu=\frac{n}{2pn+1}\,.
\end{equation}
Similarly, choosing \(m_1=2r\) and \(m_i=2\) gives the bosonic Jain sequence
\begin{equation}
    \nu=\frac{n}{(2r-1)n+1}\,.
\end{equation}

\subsection{$\Sp(4,\bZ)$ and Haldane-Halperin hierarchy}
The \(\U(1)\times \U(1)\) case describes a bilayer quantum Hall system with independent charge conservation in the two layers. We use the layer basis \(A_1,A_2\) for compact gauge fields, and introduce the charge and pseudospin combinations
\begin{equation}
    A^c = (A_1+A_2)/2 \,,\qquad A^s = (A_1-A_2)/2\,.
\end{equation}
The inverse relations are
\[
A_1=A^c+A^s,\qquad A_2=A^c-A^s \,.
\]
Consequently the charge vectors transform as:
\[
t^c=t^1+t^2,\qquad t^s=t^1-t^2 \,.
\]
Thus the two electron species have charges \((Q_c,Q_s)=(1,1)\) and
\((1,-1)\), respectively. 
These combinations are useful for charge assignments and response coefficients. Since this change of basis is not unimodular, the \(K\)-matrices below are still written in the layer basis.

The parent bilayer states are the Halperin \((m_1,m_2,n)\) states,
\begin{equation}
    K=\begin{pmatrix}
        m_1& n \\
        n & m_2
    \end{pmatrix},\qquad  t^c = (1,1),\qquad t^s = (1,-1)\,.
\label{eq:Halperin}
\end{equation}
In the continuum Galilean-invariant setting discussed above, their total filling equals the charge-sector Hall conductivity,
\begin{equation}
    \nu=\sigma_{cc}=t^{c\,T}K^{-1}t^c
    =\frac{m_1+m_2-2n}{m_1m_2-n^2}\,.
\end{equation}
We mostly focus on the layer-exchange symmetric case \(m_1=m_2=m\), for which
\begin{equation}
    \nu=\frac{2}{m+n}\,.
\end{equation}
Another important case is the \(\SU(2)\)-compatible, or spin-singlet, condition \(n=m-1\), \cite{balatsky1991singlet,milovanovic1997invariant}. With the normalization \(t^s=(1,-1)\), the \(\U(1)_s\) pseudospin charge of every quasiparticle is then integral, although the topological spins can still be fractional. Therefore the $U(1)_s$ can be enhanced to $SU(2)$, as $SU(2)$ doesn't have non-trivial fractionalization class. The filling becomes
\begin{equation}
    \nu=\frac{2}{2m-1}\,.
\end{equation}

As in the single-\(\U(1)\) construction, a hierarchy step is implemented by introducing auxiliary backgrounds through \(A_i\to A_i-B_i\) and then applying \(ST^{p_i}\) to the appropriate \(B_i\). This describes a daughter state formed from excess quasiparticles or quasiholes. Acting only on \(B_1\) gives
\begin{equation}
\label{eq:onestep}
    \frac{1}{4\pi}K_{IJ}a^I d a^J
    + \frac{1}{2\pi}a^1 d(A_1-b_1)
    + \frac{1}{2\pi}a^2 dA_2
    +\frac{p_1}{4\pi}b_1 db_1
    +\frac{1}{2\pi}b_1 dC_1 \,,
\end{equation}
which, after setting the auxiliary background \(C_1\) to zero, is described by
\begin{equation}
    K= \begin{pmatrix}
        m_1& n & -1\\
        n & m_2& 0\\
        -1& 0&p_1
    \end{pmatrix},\qquad  t^c = (1,1,0),\qquad t^s = (1,-1,0)\,.
    \label{eq:onestep-K}
\end{equation}
To preserve layer exchange, we act on both auxiliary backgrounds \(B_1,B_2\) with the diagonal gauging operation \(S_d\), followed by a layer-exchange-invariant \(T_B\) transformation, with
\begin{equation}
    T_B=\begin{pmatrix}
        p&q\\ q&p
    \end{pmatrix}.
\end{equation}
The resulting effective theory is
\begin{equation}
    K= \begin{pmatrix}
        m& n & -1&0\\
        n & m& 0&-1\\
        -1& 0&p &q \\
        0& -1 &q &p
    \end{pmatrix},\qquad  t^c = (1,1,0,0),\qquad t^s = (1,-1,0,0)\,.
\end{equation}
This describes a layer-exchange symmetric Abelian state with total filling
\begin{equation}
    \nu=\frac{2 (p+q)}{(m+n) (p+q)-1}\,.
\end{equation}
In the layer-even and layer-odd basis, equivalently the charge and pseudospin sectors, the determinant factorizes as
\begin{equation}
    \det K=
    \big((m+n)(p+q)-1\big)\big((m-n)(p-q)-1\big)\,.
\end{equation}
A fully gapped Abelian topological order therefore requires both factors to be nonzero. The choice \(q=0,\ p=2\) corresponds to decoupled \((220)\)-type daughter condensates and gives the Abelian intralayer-flux branch of the bilayer composite-fermion hierarchy \cite{scarola2001phase}.

For the higher layer-exchange-symmetric hierarchy, we consider the same additional diagonal block $\begin{pmatrix}p&q \\q&p\end{pmatrix}$, define
\begin{equation}
    \alpha=m+n,\qquad \beta=p+q \,.
\end{equation}
Let \(s=1\) denote the parent Halperin state. The total filling of the \(s\)-th hierarchy state is controlled by the layer-even continued fraction
\begin{equation}
    \nu_s
    =
    \frac{2}{\alpha-\cfrac{1}{\beta-\cfrac{1}{\beta-\cdots-\cfrac{1}{\beta}}}} \,,
\end{equation}
where the denominator contains \(s-1\) copies of \(\beta\). Thus \(\nu_1=2/\alpha\) and \(\nu_2=2/(\alpha-1/\beta)\). It is useful to define
\begin{equation}
    F_0=0,\qquad F_1=1,\qquad F_{s+1}=\beta F_s-F_{s-1}\,.
\end{equation}
Then
\begin{equation}
    \nu_s=\frac{2F_s}{\alpha F_s-F_{s-1}} \,.
\end{equation}
Equivalently, for $\beta^2\neq 4$,
\begin{equation}
    F_s=\frac{\lambda_+^{\,s}-\lambda_-^{\,s}}{\lambda_+-\lambda_-}\,,
    \qquad
    \lambda_\pm=\frac{\beta\pm\sqrt{\beta^2-4}}{2}\,,
\end{equation}
and hence
\begin{equation}
    \nu_s=
    \frac{2\left(\lambda_+^{\,s}-\lambda_-^{\,s}\right)}
    {\alpha\left(\lambda_+^{\,s}-\lambda_-^{\,s}\right)
    -\left(\lambda_+^{\,s-1}-\lambda_-^{\,s-1}\right)} \,.
\end{equation}
For the degenerate case $\beta=2$, one has $F_s=s$, so
\begin{equation}
    \nu_s=\frac{2s}{(\alpha-1)s+1}
    =\frac{2s}{(m+n-1)s+1}\,.
\end{equation}

The \(\SU(2)\)-compatible, or spin-singlet, branch is obtained by imposing \(n=m-1\) in the parent state and \(q=p-2\) in the daughter condensate. The one-step hierarchy matrix then takes the form
\begin{equation}
    K= \begin{pmatrix}
        m& m-1 & -1&0\\
        m-1 & m& 0&-1\\
        -1& 0&p &p-2 \\
        0& -1 &p-2 &p
    \end{pmatrix}\,,\qquad  t^c = (1,1,0,0)\,,\qquad t^s = (1,-1,0,0)\,.
\end{equation}
Its filling is
\begin{equation}
    \nu=\frac{4 (p-1)}{4 m (p-1)-2 p+1}\,.
\end{equation}
For the higher spin-singlet hierarchy, the layer-even parameters are
\begin{equation}
    \alpha=2m-1\,,\qquad \beta=2(p-1)\,,
\end{equation}
and hence
\begin{equation}
    \nu_s=\frac{2F_s}{(2m-1)F_s-F_{s-1}}\,,
    \qquad F_{s+1}=2(p-1)F_s-F_{s-1}\,.
\end{equation}
For \(p=2\), this reduces to
\begin{equation}
    \nu_s=\frac{2s}{(2m-2)s+1}\,.
\end{equation}

For example, for $(332)+r\times (pp,p-2)$, the total fillings $\nu_c$ are,
\begin{equation}
    \begin{array}{c|ccccc}
 & r=0 & r=1 & r=2 & r=3 &r\\ \hline
 (332),(220),\; \nu_c
& \frac{2}{5} & \frac{4}{9} & \frac{6}{13} & \frac{8}{17} & \frac{2(r+1)}{4r+5}\\\hline
(332),(00,-2),\; \nu_c
& \frac{2}{5} & \frac{4}{11} & \frac{6}{17} & \frac{8}{23} & \frac{2(r+1)}{6r+5}\\\hline
(332),(442),\; \nu_c
& \frac{2}{5} & \frac{12}{29} & \frac{70}{169} & \frac{408}{985} & ...
\end{array}
\end{equation}
where $(332)+r\times(220)$ is the ordinary Jain sequence of the spin-singlet state $(332)$, while the others are generalizations beyond Jain states.

\subsection{Application to bilayer FQH}
The recent bilayer graphene experiment of Ref.~\cite{Nguyen:2024jip} maps the bilayer FQH phase diagram as a function of the independently tunable layer fillings \(\nu_1\) and \(\nu_2\). On the equal-density line \(\nu_1=\nu_2\), the high-field, weak-interlayer-correlation regime exhibits the ordinary Jain sequence \(\nu_i=N/(2N+1)\); in our hierarchy notation this is the decoupled \((330)\)-type branch. At lower fields, where interlayer correlations are stronger, the observed insulating states are consistent with two-component FQH states with interlayer flux attachment, which are organized below as \((331)\)- and \((332)\)-type branches. The experimental data also show enhanced counterflow conductance and drag near \(\nu_1=\nu_2=3/8\), outside the integer-filled Jain sequence. One possible Abelian FQH state at this filling is given by composite fermions \({}^{2}_{0}\mathrm{CF}\), where we use \({}^{u}_{l}\mathrm{CF}\) to denote a composite fermion with \(u\) intralayer and \(l\) interlayer flux quanta attached. Each layer fills \(\nu_i^*=3/2=1+1/2\) \(\Lambda\)-levels, and the two half-filled \(\Lambda\)-levels can be modeled to form an interlayer exciton condensate, corresponding to the Halperin \((111)\) state. Analogously, \(5/12\) corresponds to the \(\nu_i^*=5/2=2+1/2\) case.

We now compare these observations with our layer-symmetric hierarchy construction. The ordinary Jain hierarchy can be reinterpreted as built from an \((mmn)\) parent uses only \((220)\) daughter blocks,
\begin{equation}
    (mmn),\quad (mmn)+(220),\quad (mmn)+2(220),\quad \ldots \,.
\end{equation}
where $+n(220)$ means adding $n$ copies of the $(220)$ block in the $K$-matrix together with coupling matrix,
\begin{equation}
    K_{11} =\begin{pmatrix}
        m&n\\n&m
    \end{pmatrix}  \,,\quad K_{u,u} = \begin{pmatrix}
        2&0\\0&2
    \end{pmatrix}\,,\quad K_{u,u+1}=K_{u+1,u}=\begin{pmatrix}
        -1&0\\0&-1
    \end{pmatrix}\,,
\end{equation}
where the subscripts of $K$ denote the position of the $2\times 2$ blocks. To reach the fractions beyond the Jain-like sequence for the layer symmetric case, we allow the final daughter block to be the interlayer-correlated \((221)\) block and alternate it with the Jain steps. This is allowed since we only require the extra quasi-particle/quasi-hole to form a bilayer bosonic FQH state and $(221)$ has lower filling $2/3$ compared with $(220)$ whose filling is $1$. To be specific, the sequence is given by,
\begin{equation}
    \begin{aligned}
    (mmn)
    &\to
    (mmn)+(221)
    \to
    (mmn)+(220)\\
    &\to
    (mmn)+(220)+(221)
    \to
    (mmn)+2(220)
    \to \cdots \,.
    \end{aligned}
\end{equation}
where the off-diagonal coupling matrix added implicitly. The index \(s\) labels the level in the alternating sequence, related to the number \(r\) of \((220)\) blocks by \(s=2r+1\) for the odd (ordinary Jain) entries and \(s=2r+2\) for the even entries. Thus the odd entries are ordinary Jain states \((mmn)+r(220)\), while the even entries are the new states \((mmn)+r(220)+(221)\), with \(r=0,1,\ldots\). This alternating construction gives the following total fillings \(\nu_{\rm tot}=\nu_1+\nu_2\):
\begin{equation}
\begin{array}{c|ccccccc}
 & s=1 & s=2 & s=3 & s=4 & s=5 & s=6 & s \\ \hline
(330) & \frac{2}{3} & \frac{3}{4} & \frac{4}{5} & \frac{5}{6} & \frac{6}{7} & \frac{7}{8}& \frac{s+1}{s+2} \\ \hline
(331) & \frac{1}{2} & \frac{6}{11} & \frac{4}{7} & \frac{10}{17} & \frac{3}{5} & \frac{14}{23}& \frac{2s+2}{3s+5} \\ \hline
(332) & \frac{2}{5} & \mathrm{EC},\,\frac{3}{7} &\frac{4}{9} &
\mathrm{EC},\,\frac{5}{11} & \frac{6}{13} &
\mathrm{EC},\,\frac{7}{15} & \frac{s+1}{2s+3}
\end{array}
\end{equation}

Here \(\mathrm{EC}\) denotes a singular \(K\)-matrix with \(\det K=0\), which contains a gapless mode and may be interpreted as an exciton condensation phase, since the null vector $t^0$ of $K$ has non-zero overlap with $t^s$, and gives the EC response theory. Although \(K^{-1}\) does not exist, the charge filling is still well-defined when \(Kx=t^c\) has a solution, with \(t^c=(1,1,0,\ldots)^T\). In that case \(\nu_c=t^{c\,T}x\) is independent of the choice of solution, and we display the singular entries as \(\mathrm{EC},\nu_c\). For the \((330)\) row, the corresponding equal-layer fillings are
\begin{equation}
    \begin{array}{c|cccccc}
 & s=1 & s=2 & s=3 & s=4 & s=5 & s=6 \\ \hline
(330),\; \nu_1=\nu_2
& \frac{1}{3} & \frac{3}{8} & \frac{2}{5} & \frac{5}{12} & \frac{3}{7} & \frac{7}{16}
\end{array}
\end{equation}
Here the \(s=1,3,5,\ldots\) entries are the ordinary Jain states \((330)+r(220)\), while the \(s=2,4,6,\ldots\) entries are the new states \((330)+r(220)+(221)\). Thus \(3/8\) and \(5/12\) arise from the final \((221)\) block and lie beyond the ordinary Jain sequence. The topological order is equivalent to the abelian $\mathbb{Z}_{8(r+2)}$ with $c_- = 2r+4$ and minimal charge $\frac{1}{4(r+2)}$.

For the \((330)\) parent, the ordinary branch \((330)+r(220)\) has
\begin{equation}
    \nu_{\rm tot}^{(220)}(r)=\frac{2(r+1)}{2r+3},\qquad
    \nu_1=\nu_2=\frac{r+1}{2r+3}.
\end{equation}
The new branch \((330)+r(220)+(221)\) has
\begin{equation}
    \nu_{\rm tot}^{(221)}(r)
    =
    \frac{2}{3-\cfrac{1}{2-\cfrac{1}{\cdots-\cfrac{1}{3}}}}
    =
    \frac{2r+3}{2r+4}\,,\qquad
    \nu_1=\nu_2=\frac{2r+3}{4r+8}\,,
\end{equation}
where the continued fraction contains \(r\) intermediate \((220)\) blocks before the final \((221)\) block. The \(r=0\) member gives \(3/8+3/8\), whose candidate Abelian hierarchy matrix is
\begin{equation}
\label{eq:threequarter}
    K^4= \begin{pmatrix}
        3& 0 & -1&0\\
        0 & 3& 0&-1\\
        -1& 0&2 &1 \\
        0& -1 &1 &2
    \end{pmatrix},\qquad  t^c = (1,1,0,0)\,,\qquad t^s = (1,-1,0,0)\,.
\end{equation}
Such Abelian FQH for \(3/8+3/8\) is equivalent to spin Abelian topological order with bosonic part $\bZ_{16}^{(15)}$, whose minimal anyon has topological spin $15/32$ and $U(1)^c$ charge $1/8$. There is also the non-Abelian proposal for such filling with distinct anyon contents \cite{2017jain-nonabelian}. 

The \(r=1\) member gives \(5/12+5/12\), represented by
\begin{equation}
\label{eq:fivesixth}
    K^6= \left(
\begin{array}{cccccc}
 3 & 0 & -1 & 0 & 0 & 0 \\
 0 & 3 & 0 & -1 & 0 & 0 \\
 -1 & 0 & 2 & 0 & -1 & 0 \\
 0 & -1 & 0 & 2 & 0 & -1 \\
 0 & 0 & -1 & 0 & 2 & 1 \\
 0 & 0 & 0 & -1 & 1 & 2
\end{array}
\right),\qquad t^c = (1,1,0,0,0,0)\,,\qquad t^s=(1,-1,0,0,0,0)\,.
\end{equation}
These matrices implement the final-\((221)\) branch of the alternating hierarchy. The energetic preference between candidate paths requires microscopic input, so they should be viewed as candidate Abelian effective descriptions of the equal-layer fractions discussed in Ref.~\cite{Nguyen:2024jip}.

These hierarchy matrices are equivalent to the Abelian composite-fermion matrices by integral changes of anyon basis. For \(3/8+3/8\), the Abelian composite-fermion matrix may be written as
\begin{equation}
    K^4_{\mathrm{CF}}= \begin{pmatrix}
        3& 2 & 0&0\\
        2 & 3& 0&1\\
        0& 0&3 &2 \\
        0& 1 &2 &3
    \end{pmatrix}\,,\qquad
t^c_{\mathrm{CF}} = (1,1,1,1)\,,\qquad t^s_{\mathrm{CF}} = (1,1,-1,-1)\,.
\end{equation}
The basis is ordered as \(\psi_{\ell,\alpha}=(\psi_{1,0},\psi_{1,1},\psi_{2,0},\psi_{2,1})\), where \(\ell\) labels the layer and \(\alpha\) labels the \(\Lambda\)-level. The relation to Eq.~\eqref{eq:threequarter} is
\begin{equation}
    W_4=
    \begin{pmatrix}
        1&1&0&0\\
        0&0&1&1\\
        0&1&0&0\\
        0&0&0&1
    \end{pmatrix}\in\GL(4,\bZ)\,,\qquad
W_4^\intercal K^4 W_4=K^4_{\mathrm{CF}}\,,\qquad W_4^\intercal t^{c/s}=t^{c/s}_{\mathrm{CF}} \,.
\end{equation}
For \(5/12+5/12\), the composite-fermion matrix is given by,
\begin{equation}
K^6_{\mathrm{CF}}
=
\begin{pmatrix}
3 & 2 & 2 & 0 & 0 & 0 \\
2 & 3 & 2 & 0 & 0 & 0 \\
2 & 2 & 3 & 0 & 0 & 1 \\
0 & 0 & 0 & 3 & 2 & 2 \\
0 & 0 & 0 & 2 & 3 & 2 \\
0 & 0 & 1 & 2 & 2 & 3
\end{pmatrix},
\qquad
t^c_{\mathrm{CF}}
=
(1,1,1,1,1,1),
\qquad
t^s_{\mathrm{CF}}
=
(1,1,1,-1,-1,-1).
\end{equation}
and the analogous relation for \(K^6\) in Eq.~\eqref{eq:fivesixth} is
\begin{equation}
    W_6=
    \begin{pmatrix}
        1&1&1&0&0&0\\
        0&0&0&1&1&1\\
        0&1&1&0&0&0\\
        0&0&0&0&1&1\\
        0&0&1&0&0&0\\
        0&0&0&0&0&1
    \end{pmatrix}\in\GL(6,\bZ),\qquad
W_6^\intercal K^6 W_6=K^6_{\mathrm{CF}}\,,\qquad W_6^\intercal t^{c/s}=t^{c/s}_{\mathrm{CF}} .
\end{equation}
Both \(W_4\) and \(W_6\) have determinant \(-1\). They preserve the anyon lattice, braiding data, and charge assignments. Hence the hierarchy construction with a final \((221)\) block and the Abelian \({}^{2}_{0}\mathrm{CF}\) construction describe the same Abelian topological order.

More generally, the \((33n)\) rows can be viewed as composite-fermion states with two intralayer and \(n\) interlayer flux quanta attached. Thus \(n=0,1,2\) gives the \((330)\), \((331)\), and \((332)\) hierarchy branches. For \(n=2\), the even hierarchy levels in the table are singular and describe the exciton-condensate branch, where the pseudospin \(U(1)\) symmetry is spontaneously broken.

\begin{figure}[t]
    \centering
    \includegraphics[width=0.9\textwidth]{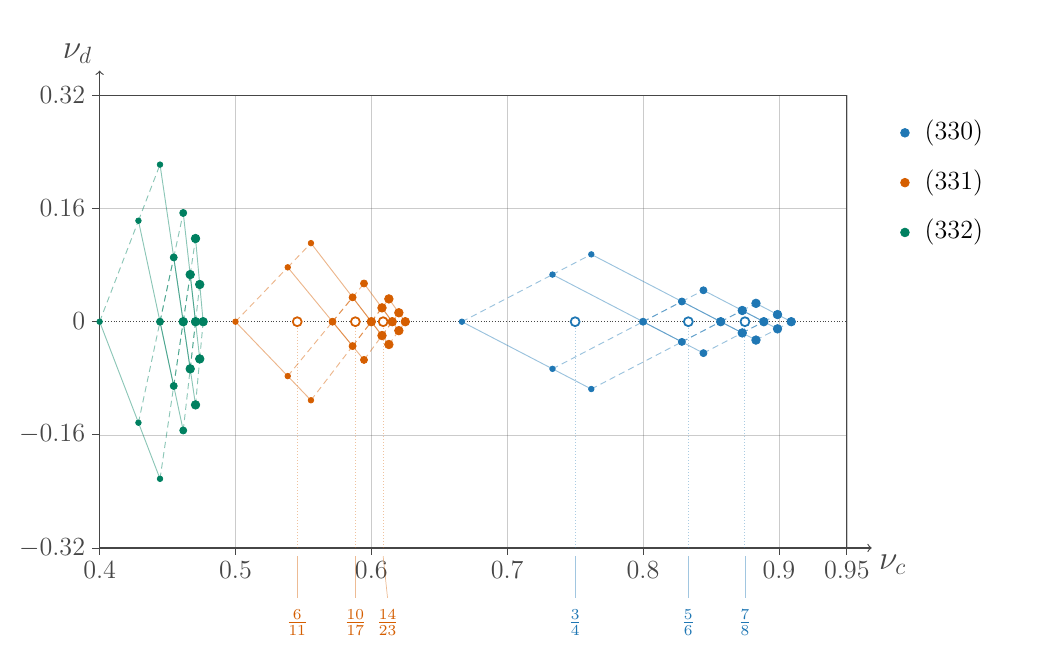}
    \caption{Layer-imbalanced hierarchy grid in the \((\nu_c,\nu_d)\) plane for \((33n)+r(220)+p[2]_u+q[2]_\ell\), with \(n,r=0,1,2\) and \(p,q=0,1,2\). The horizontal coordinate is the total charge response \(\nu_c\), and the vertical coordinate is the layer-imbalance response \(\nu_d=\nu_u-\nu_\ell\). Solid lines fix the upper-layer Jain level, dashed lines fix the lower-layer Jain level, and the horizontal center line is \(\nu_d=0\). Solid circles denote the ordinary Jain-sequence grid, while open circles mark the layer-symmetric final-\((221)\) branch \((33n)+r(220)+(221)\), with \(r=0,1,2\), placed on the equal-layer line.}
    \label{fig:imbalanced-grid}
\end{figure}

\paragraph{Layer-imbalanced case.}
Away from the equal-density line, the two layers may lie at different Jain levels. Starting from an \((mmn)\) parent, the corresponding asymmetric hierarchy is obtained by attaching independent one-component Jain daughter chains to the upper and lower endpoints,
\begin{equation}
    (mmn)+p[2]_u+q[2]_\ell\,,\qquad p,q\in\mathbb Z_{\ge 0}\,.
\end{equation}
Here \([2]_u\) and \([2]_\ell\) denote \(K=2\) scalar blocks attached to the upper and lower chains, respectively. For nonsingular \(K\), the total charge response is
\begin{equation}
    \nu_c=t^{c\,T}K^{-1}t^c\,,
\end{equation}
And we define the upper and lower filling by,
\begin{equation}
    \nu_u = t^{u\,T} K^{-1}t^c,\quad \nu_\ell = t^{\ell\,T} K^{-1}t^c\,.
\end{equation}
where $t^u=(1,0,\cdots)$ and $t^\ell=(0,1,\cdots)$. Therefore, the layer-imbalance response is, 
\begin{equation}
    \nu_d = \nu_u-\nu_\ell = t^{u\,T}K^{-1}t^u-t^{\ell\,T}K^{-1}t^\ell\,.
\end{equation}
Thus \(\nu_d=\nu_1-\nu_2\) matches the experimental layer notation, while \(\nu_c\) is still computed from the total charge vector rather than by adding the two diagonal layer responses. This gives a two-dimensional grid of hierarchy states, shown in Fig.~\ref{fig:imbalanced-grid} for the \((330)\), \((331)\), and \((332)\) parents.

\begin{figure}[t]
    \centering
    \includegraphics[width=0.9\textwidth]{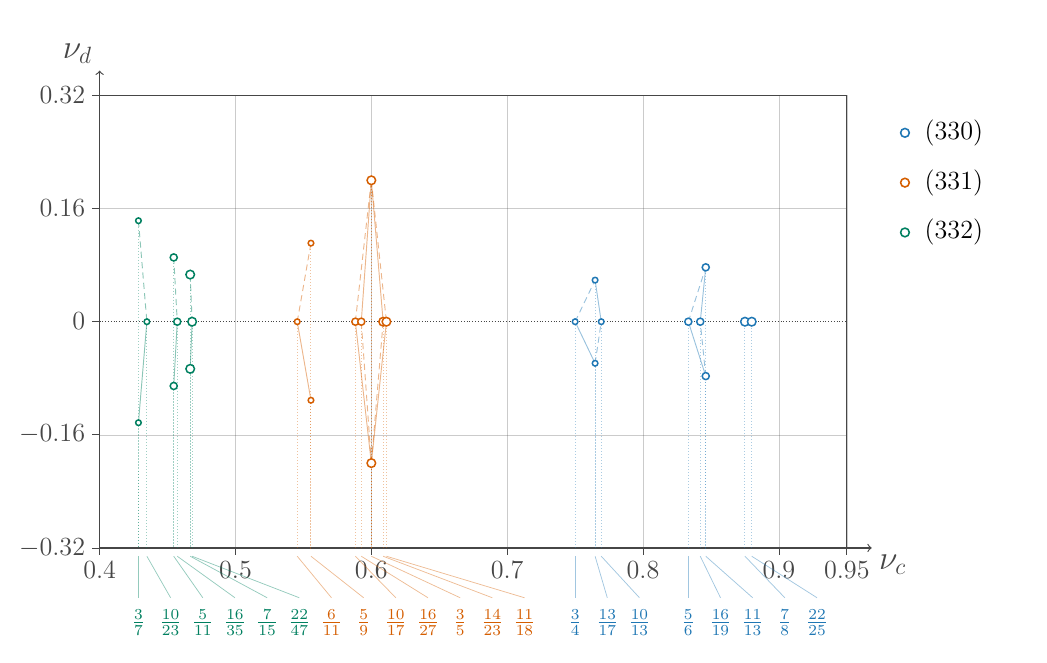}
    \caption{Layer-imbalanced plot in the \((\nu_c,\nu_d)\) plane for \((33n)+r(220)+(221)+p[2]_u+q[2]_\ell\), with \(n,r=0,1,2\) and \(p,q=0,1\). The colors label the parent \((330)\), \((331)\), or \((332)\). Within each fixed \(r\) slice, solid lines fix the upper scalar level \(p\), dashed lines fix the lower scalar level \(q\). Singular entries for which \(\nu_d\) is not defined are omitted.}
    \label{fig:balanced-221-imbalanced-grid}
\end{figure}

A second imbalanced extension keeps the final-\((221)\) branch and then attaches independent scalar Jain chains to the two endpoints,
\begin{equation}
    (mmn)+r(220)+(221)+p[2]_u+q[2]_\ell,\qquad r,p,q\in\mathbb Z_{\ge 0}\,.
\end{equation}
The resulting plot for \(r=0,1,2\) and \(p,q=0,1\) is shown in Fig.~\ref{fig:balanced-221-imbalanced-grid}, and singular entries with undefined \(\nu_d\) are omitted.

\subsection{A note on U(1) versus spin-$c$ connections}

A standard $U(1)$ gauge field obeys the familiar flux quantization:
\begin{equation}
    \oint_{\mathcal{C}} \frac{F}{2\pi} \in \mathbb{Z}\,,
\end{equation}
where $\mathcal{C}$ is a homologically nontrivial two-cycle. We often instead couple to a $\text{Spin}_c$ connection, which obeys:
\begin{equation}
    \oint_{\mathcal{C}} \frac{F}{2\pi} = \frac{1}{2}\oint_{\mathcal{C}} w_2 \mod \mathbb{Z} \, , 
\end{equation}
where $w_2$ is the second Stiefel-Whitney class of the tangent bundle of spacetime. In this subsection, we discuss the physics of $\text{Spin}_c$ connections and the constraints on $K$-matrices implied by coupling to them.
\subsubsection{General setup}

Following \cite{Seiberg:2016rsg,Seiberg:2016gmd} we want to understand what additional constraints are imposed on our system by spin-charge relations. Whenever a system has the property that all fermions carry odd charge while all bosons carry even charge (as is true in systems made of protons and electrons), the way the partition function depends on $U(1)$ backgrounds and spin structure is constrained in a way that is best packaged by introducing a spin-$c$ connection. On a spin 3-manifold -- in three dimensions every orientable spin-$c$ manifold is automatically spin, so the spin-$c$ data is nothing but a U(1) connection together with a choice of spin structure -- a theory obeying spin-charge depends on the U(1) background $A$ and the spin structure only through a combination that is invariant under a \emph{simultaneous} shift of the two.

Concretely, insert a flat Wilson line of holonomy $-1$ (a ``$\pi$'' background) around an arbitrary closed cycle. Every field of odd charge -- the fermions -- picks up an extra sign around that cycle, while even-charge fields -- the bosons -- are unaffected. This is exactly the effect of changing the spin structure along that cycle. The two operations therefore cancel: the shift of the U(1) background can be undone by a compensating change of spin structure, so the partition function depends only on the spin-$c$ combination. This invariance is precisely what the spin-charge relation guarantees.

Spin-charge implies certain constraints on a $K$-matrix theory of dynamical U(1) fields $b_i$ coupled to a single spin-$c$ background $A$:
\begin{equation}
\mathcal{L}
= \frac{k_{ij}}{4\pi}\, b_i\, db_j
+ \frac{q_i}{2\pi}\, b_i\, dA
+ \frac{k}{4\pi}\, A\, dA
+ (2k + 16n)\,\mathrm{CS}_g \, ,
\label{eq:SSWW-K}
\end{equation}
with $k_{ij}, q_i, k, n \in \mathbb{Z}$ and the spin-charge constraint
\begin{equation}
k_{ii} \equiv q_i \pmod{2} \, .
\label{eq:SSWW-diag}
\end{equation}
The constraint applies only to the \emph{diagonal} entries of $k$; the off-diagonals are $U(1)$-$U(1)$ BF couplings and may be any integer.

In our setup the individual layer symmetries do not separately obey the
spin-charge relation for the full bilayer theory: the electron in layer 1 is
neutral under \(U(1)_2\), and the electron in layer 2 is neutral under
\(U(1)_1\). The diagonal charge symmetry does obey spin-charge, since both
electron species have
$Q_c=Q_1+Q_2=1$.
The pseudospin symmetry has the same mod-two charge assignment, since
$Q_s=Q_1-Q_2$ with $Q_c-Q_s=2Q_2$.
Thus every local excitation has
\begin{equation}
Q_c\equiv Q_s\pmod 2 \,.
\end{equation}
Consequently imposing the spin-charge constraint using \(A^c\) or using
\(A^s\) gives the same mod-two condition on the \(K\)-matrix.

Importantly, the dynamical fields stay in the layer basis $(a_1, a_2)$, where individually they do not always satisfy a spin-charge relation. They must therefore be summed over as ordinary U(1) connections, not as spin-$c$ connections. This is also the standard FQH effective-theory choice.

\subsubsection{Constraints for the bilayer system}

We impose the spin-charge constraint for the single spin-$c$ background $A_c$. The dynamical fields appearing in the $K$-matrix -- the layer fields $a_1, a_2$ and every auxiliary $b_i$ introduced by an $ST^{p_i}$ step -- are ordinary U(1). The rule \eqref{eq:SSWW-diag} then imposes the single condition
\begin{equation}
\boxed{\;
K_{ii} \equiv t^c_i \pmod 2, \qquad
K_{ij}\ (i\neq j) \in \mathbb{Z}\ \text{unrestricted,}
\;}
\label{eq:setupB-constraint}
\end{equation}
where $t^c_i$ is the charge of $b_i$ under $U(1)_c$. The off-diagonals are unconstrained because they are U(1)--U(1) BF couplings. Note that it makes no difference whether we pick $A_c$ or $A_s$ as the spin-$c$ background: choosing $A_s$ gives $K_{ii} \equiv t^s_i$, which is the \emph{same} condition. The point is a statement about charges, not connections. The two diagonal charges differ by the  $Q_c - Q_s = 2\,Q_2$ charge, and every field carries \emph{even} charge there, since it is twice an integer layer-2 charge:
\begin{align}
   t^c_i - t^s_i = 2 t_i^2 = 0\pmod 2 \,. 
\end{align}
Hence $t^c_i \equiv t^s_i \pmod 2$ and the constraints are identical.

It is straightforward to check that the constraint \eqref{eq:setupB-constraint} is indeed obeyed by all the $K$-matrices in this section. For the Halperin state of \eqref{eq:Halperin} with $t^c = (1,1)$ the constraint demands that the diagonal entries $m_1, m_2$ are odd while the off-diagonal $n$ is unrestricted, which is the correct constraint for all fermionic Halperin states. For the one-step hierarchy on $B_1$ from \eqref{eq:onestep} we have $t^c = (1,1,0),\ t^s = (1,-1,0)$, so we need $m_1, m_2$ to be odd and $p_1$ to be even whereas the off-diagonals $n, -1, 0$ are unrestricted. $p_1$ even is telling us that we did not introduce any new fermions when introducing the extra dynamical fields; only bosons ``condensed" when going down the hierarchy. This pattern repeats when going further down the hierarchy: the diagonal entries corresponding to the "electron rows" in the $K$-matrix where the entries in $t^c$ are 1 have to be odd, whereas all the other entries have to be even. Concretely, for the $\nu_{\rm tot} = 3/4$ bilayer of \eqref{eq:threequarter} with $t^c = (1,1,0,0)$ the diagonal entries $3,3,2,2$ are indeed odd for the first two entries and even in all the others, and similarly for the $\nu_{\rm tot} = 5/6$ bilayer of \eqref{eq:fivesixth} with $t^c = (1,1,0,0,0,0)$ and diagonal entries $3,3,2,2,2,2$.

\section{Conclusion and future directions}
\label{sec:conclusion}
In this work, we formulated the \(\Sp(4,\mathbb{Z})\) action on 3d theories with \(U(1)\times U(1)\) global symmetry and used it to organize a web of distinct infrared phases. The construction follows from the bulk-boundary correspondence: electromagnetic duality transformations of a 4d \(U(1)^2\) gauge theory induce theory-generating operations on its 3d boundary. Beyond transformations inherited from the individual \(U(1)\) factors, we identified an intrinsically two-component order-five element \(g_{5}\in \Sp(4,\mathbb{Z})\). Its bulk action satisfies \(g_{5}^{5}=I\), while the induced boundary operation closes only up to stacking with a decoupled \(U(1)_{1}\) invertible phase. This provides a projective realization of the order-five duality and a boundary signature of a mixed duality-gravity anomaly. More generally, the \(\Sp(4,\mathbb{Z})\) operations relate symmetric gapped phases with Abelian topological order, invertible phases, and symmetry-breaking phases with Nambu--Goldstone modes. In the \(K\)-matrix description, the diagonal and off-diagonal \(S\)- and \(T\)-type operations provide a systematic framework for tracking the topological order and \(U(1)\times U(1)\) symmetry-fractionalization data. The construction extends directly to theories with \(U(1)^n\) symmetry, for which the corresponding transformations are governed by \(\Sp(2n,\mathbb{Z})\).

We applied this framework to bilayer fractional quantum Hall systems, where the two conserved layer charges may be reorganized into the electromagnetic \(U(1)_c\) and pseudospin \(U(1)_s\) symmetries. We first reinterpret the conventional hierarchy construction for single-component fractional quantum Hall states in terms of \(\SL(2,\mathbb{Z})\) transformations. We then generalize this construction using \(\Sp(4,\mathbb{Z})\) operations, which naturally generate bilayer Haldane-Halperin hierarchies, including branches beyond the ordinary Jain-type sequence. In these constructions, the additional quasiparticles or quasiholes form bilayer bosonic fractional quantum Hall states. This yields layer-exchange-symmetric and \(SU(2)\)-compatible spin-singlet branches, as well as candidate Abelian states at even-denominator equal-layer fillings beyond the ordinary Jain sequence, including \(3/8+3/8\) and \(5/12+5/12\). Their equivalence to Abelian composite-fermion descriptions under integral changes of anyon basis shows that the \(\Sp(4,\mathbb{Z})\) construction provides a unified bridge between 3d duality operations, symmetry-enriched topological order, and experimentally relevant bilayer quantum Hall hierarchies.

A further direction is to extend the \(\Sp(4,\mathbb{Z})\) construction to non-Abelian fractional quantum Hall phases. Familiar examples such as the Moore--Read and Read--Rezayi states admit a charged--neutral description involving a non-Abelian neutral sector and an Abelian \(U(1)\) charge sector, together with a selection rule or quotient that identifies the physical local excitations \cite{MooreRead1991,ReadRezayi1999}. More generally, a non-Abelian hierarchy state can be constructed by stacking a non-Abelian topological order \(\mathcal{C}\) with an Abelian \(K\)-matrix theory \(\mathcal{A}_{K}\) and condensing a suitable algebra of bosonic anyons,
\[
    \mathcal{C}_{\mathrm{FQH}}
    =
    \bigl(\mathcal{C}\boxtimes\mathcal{A}_{K}\bigr) /\mathcal{B},
\]
as in the categorical formulation of anyon condensation and its application to non-Abelian quantum Hall hierarchies \cite{Kong2014,BondersonSlingerland2008,Zhang:2024bye,
BarkeshliWen2011Bilayer}. In such a presentation, the \(\Sp(4,\mathbb{Z})\) operations act naturally on the Abelian sector and its \(U(1)\times U(1)\) symmetry-fractionalization data, while the condensable algebra must be mapped consistently to the transformed theory. More general transformations may additionally act on the non-Abelian sector. This suggests a theory-generating web relating distinct \(U(1)\times U(1)\)-symmetric non-Abelian fractional quantum Hall phases, including two-component and bilayer states \cite{BarkeshliWen2010TwoComponent,BarkeshliWen2011Bilayer}.

Our analysis has focused mainly on how the \(\Sp(4,\mathbb{Z})\) operations relate infrared phases. A natural next step is to construct matter-coupled realizations and determine how the intervening phase transitions, critical theories, and their relevant deformations are mapped under the same operations \cite{Witten:2003ya,Senthil:2018cru,Wang:2017txt}. Of particular interest is the order-five transformation \(g_5\), whose projective boundary action is related to the duality--gravity anomaly. It would be interesting to determine whether the phases in an order-five orbit can meet at a \(g_5\)-invariant multicritical point, with the relevant deformations and adjacent phases cyclically permuted by the duality, generalizing known examples of self-dual criticality in gauge theories with matter \cite{Xu:2015lxa,Cheng:2016pdn,Jian:2017chw,Lu:2021ucu,Dumitrescu:2026vre}. More generally, the construction can be extended to theories with \(U(1)^n\) global symmetry and duality group \(\Sp(2n,\mathbb{Z})\). It would be useful to classify intrinsically higher-rank finite-order elements, determine the invertible phases that appear in their projective boundary relations, and clarify how these data encode the corresponding duality-gravity anomalies and their relation to higher-genus mapping class groups \cite{Seiberg:2018ntt,benson2018cohomology}.

\section*{Acknowledgments}
We are grateful to Clemens Kuhlenkamp, Jia Li, Ashvin Vishwanath, Taige Wang for helpful discussions. The work of YA, AK, and RS was supported in part by DOE grant DE-SC0022021 and by a grant from the Simons Foundation (Grant 651678, AK). D.C.L. is supported by the Simons Collaboration on Ultra-Quantum Matter, which is a grant from the Simons Foundation (grant No. 651440).

\appendix

\section{Presentation of $\Sp(4,\bZ)$}
\label{app:presentation}
We follow \cite{Behr1975} in presenting the presentation of $\Sp(4,\bZ)$. Consider the matrices 
\begin{align}
            \sR_2 &= \left(
    \begin{array}{cccc}
     1 & 1 & 0 & 0 \\
     0 & 1 & 0 & 0 \\
     0 & 0 & 1 & 0 \\
     0 & 0 & -1 & 1 \\
    \end{array}
    \right) ;\,\sT_2 = \left(
    \begin{array}{cccc}
     1 & 0 & 0 & 0 \\
     0 & 1 & 0 & 1 \\
     0 & 0 & 1 & 0 \\
     0 & 0 & 0 & 1 \\
    \end{array} \right) ;\, \sT_o =\left(
    \begin{array}{cccc}
     1 & 0 & 0 & 1 \\
     0 & 1 & 1 & 0 \\
     0 & 0 & 1 & 0 \\
     0 & 0 & 0 & 1 \\
    \end{array} \right) \, ;\nonumber\\
    \sT &= \left(
    \begin{array}{cccc}
     1 & 0 & 1 & 0 \\
     0 & 1 & 0 & 0 \\
     0 & 0 & 1 & 0 \\
     0 & 0 & 0 & 1 \\
    \end{array}
    \right) ;\,\sS_1 = \left(
    \begin{array}{cccc}
     0 & -1 & 0 & 0 \\
     1 & 0 & 0 & 0 \\
     0 & 0 & 0 & -1 \\
     0 & 0 & 1 & 0 \\
    \end{array} \right) ;\, \sS_2 =\left(
    \begin{array}{cccc}
     1 & 0 & 0 & 0 \\
     0 & 0 & 0 & -1 \\
     0 & 0 & 1 & 0 \\
     0 & 1 & 0 & 0 \\
    \end{array} \right) \, .
\end{align}
$\sR_2, \sT_o,$ and $\sT$ are the same generators as in the main text. The others can be related to the generators in the main text as 
\begin{align}
    \quad \sT_2=\sR_1\sT \sR_1,\quad \sS_1= \sS^2 \sR_1,\quad \sS_2=\sR_1 \sS^{-1} \sR_1\, .
\end{align}
Note $\sS_1^2=-1$.
At the level of the 3d lagrangian theory, we can think of the inverse T-type transformations as stacking a negative sign Chern-Simons term $T^{-1}:\cL(A_1,A_2)\mapsto\cL(A_1,A_2)-\frac{1}{4\pi}A_1dA_1$, and the inverse S-type transformations to be defined as $S^{-1}:=S^3$ as $S^4=1$. It is obvious that $R_1^{-1}=R_1$ and $R_2^{-1}: \cL(A_1,A_2) \mapsto \cL(A_1,A_2-A_1)$. The definition of the inverse of other generators follow. 

It is worth mentioning that some of the expressions below (primarily the latter with several S-Type transformations) also come with projective phases.

We can give a presentation below:
\begin{align}
\sR_2\sT_2\sR_2^{-1}\sT_2^{-1}
    &= \sT_o\,\sT\, ,\\
\sR_2\sT_o \sR_2^{-1}\sT_o^{-1}
    &= \sT^{\,2}\,,\\
\sR_2\sT\sR_2^{-1}\sT^{-1}
    &= I\,,\\
\sT_2\sT_o \sT_2^{-1}\sT_o^{-1}
    &= I\,,\\
\sT_2\sT\sT_2^{-1}\sT^{-1}
    &= I\,,\\
\sT_o\,\sT\,\sT_o^{-1}\sT^{-1}
    &= I\,,\\
\sS_1\sS_2^{2}\sS_1
    &= \sS_2^2\,,\\
\sS_2\sS_1^{2}\sS_2^{-1}
    &= \sS_1^2\,,\\
(\sS_1\sS_2)^2
    &= (\sS_2\sS_1)^2\,,\\
\sS_2^{\,4}
    &= I\,, \\
\sS_1\sT_2\sS_1^{-1}
    &= \sT\,,\\
\sS_1\sT\sS_1^{-1}
    &= \sT_2\,,\\
\sS_1\sT_o \sS_1^{-1}
    &= \sT_o^{-1}\,,\\
\sS_2\sR_2\sS_2^{-1}
    &= \sT_o\,,\\
\sS_2\sT_o\sS_2^{-1}
    &= \sR_2^{-1}\,,\\
\sS_2\sT\sS_2^{-1}
    &= \sT\,,\\
\sS_1\sR_2\sS_1\sR_2\sS_1\sR_2
    &=-I \,,\\
\sS_2\sT_2\sS_2\sT_2\sS_2\sT_2
    &= \sS_2^2\, .
\end{align}

\section{Matrix description of the web of symmetric theories}\label{app:mat}
While somewhat outside of the main development of this paper, it is useful to express the discussion of the web of $U(1)$ and $U(1)\times U(1)$ theories in terms of matrices.  These matrices extend the $K$ matrices that contain the topologically ordered part of the theory to include the charge vectors and couplings between backgrounds. 

We begin with the case of a single $U(1)$ symmetry.  As discussed in the paper, the most general Lagrangian we consider is
\begin{equation}
    \mathcal{L} = \star j a^1 +\frac{K_{IJ}}{4\pi}a^I d a^J+\frac{m}{4\pi}AdA+\frac{t}{2\pi}a^N d A
\end{equation}
and we can write it compactly and formally as
\begin{equation}
    \mathcal{L} =\begin{pmatrix}
        \cdots& a^I& \cdots &A&1
    \end{pmatrix}\left(
\begin{array}{ccccc}
  & \ldots  &  & 0 & \frac{j}{2} \\
 \ldots  & K_{IJ} & \ldots  & 0 & 0 \\
 & \ldots  &  & t & 0 \\
 0 & 0 & t & m & 0 \\
 \frac{j}{2} & 0 & 0 & 0 & 0 \\
\end{array}
\right) \begin{pmatrix}
        \cdots\\ \frac{da^I}{4\pi} \\\cdots \\\frac{dA}{4\pi}\\1
    \end{pmatrix}.
\end{equation}
Since $W^\intercal K W$ is equivalent to $K$ for $W\in GL(N,\mathbb{Z})$, 
\begin{equation}
   \left(
\begin{array}{ccccc}
  & \ldots  &  & 0 & \frac{j}{2} \\
 \ldots  & K_{IJ} & \ldots  & 0 & 0 \\
 & \ldots  &  & t & 0 \\
 0 & 0 & t & m & 0 \\
 \frac{j}{2} & 0 & 0 & 0 & 0 \\
\end{array}
\right) \simeq \left(
\begin{array}{ccccc}
  & \ldots  &  & W_{N,1}t & W_{1,1}\frac{j}{2}  \\
 \ldots  & W^\intercal K_{IJ} W & \ldots  & \cdots & \cdots \\
 & \ldots  &  & W_{N,N}t & W_{1,N}\frac{j}{2} \\
 W_{N,1} t & \cdots & W_{N,N}t & m & 0 \\
 W_{1,1}\frac{j}{2} & \cdots & W_{1,N}\frac{j}{2} & 0 & 0 \\
\end{array}
\right) 
\end{equation}
For example, $(ST)^3$ on the theory $\mathcal{L}=j\cdot A$ will give,
\begin{multline}
    \left(
\begin{array}{ccccc}
 1 & 1 & 0 & 0 & \frac{j}{2} \\
 1 & 1 & 1 & 0 & 0 \\
 0 & 1 & 1 & 1 & 0 \\
 0 & 0 & 1 & 0 & 0 \\
 \frac{j}{2} & 0 & 0 & 0 & 0 \\
\end{array}
\right) \simeq \left(
\begin{array}{ccccc}
 0 & 0 & -1 & 0 & \frac{j}{2} \\
 0 & 1 & 0 & 0 & 0 \\
 -1 & 0 & 0 & 1 & 0 \\
 0 & 0 & 1 & 0 & 0 \\
 \frac{j}{2} & 0 & 0 & 0 & 0 \\
\end{array}
\right) \\ \rightarrow j\cdot a^1 -\frac{1}{2\pi} a^1 da^3+\frac{1}{2\pi}a^3 dA +\frac{1}{4\pi} a^2 da^2
\end{multline}
It is interesting to notice that the matrix without $j$ has a non-empty kernel, indicating the reduction of gauge fields when $(ST)^{3n}$, similarly structure applies to $S^{2n}$.  Thus, the $K$-matrix contains information about what operations should yield the identity \footnote{Possibly up to a phase, as discussed above.}.

$SL(2,\mathbb{Z})$ is generated by $S$ and $T$, a general element is expressed as $S T^{m_1} S T^{m_2}\cdots S T^{m_N}$, which corresponds to the tridiagonal $K_\text{tri}$ matrix with diagonal $m_i$s and off-diagonal $1$. 

We now move to the case of $U(1) \times U(1)$.  Now the most general Lagrangian is
\begin{multline}
    \mathcal{L} = \star j a^1 + \frac{K_{IJ}}{4\pi} a^I da^J + \frac{m_1}{4\pi}A_1 dA_1 + \frac{m_2}{4\pi}A_2 dA_2\\ +\frac{n}{2\pi}A_1dA_2+ \frac{t_1}{2\pi}a^N dA_1 + \frac{t_2}{2\pi}a^N dA_2.
\end{multline}
In matrix form, this is 
\begin{equation}
    \mathcal{L} =\begin{pmatrix}
        \cdots& a^I& \cdots &A_1 & A_2&1
    \end{pmatrix}\left(
\begin{array}{cccccc}
  & \ldots  &  & 0 & 0  & \frac{j}{2} \\
 \ldots  & K_{IJ} & \ldots  & 0 & 0 & 0  \\
 & \ldots  &  & t_1 & t_2 & 0  \\
 0 & 0 & t_1 & m_1 & n & 0  \\
 0 & 0 & t_2 & n & m_2 & 0 \\
 \frac{j}{2} & 0 & 0 & 0 & 0 & 0  \\
\end{array}
\right) \begin{pmatrix}
        \cdots\\ \frac{da^I}{4\pi} \\\cdots \\\frac{dA_1}{4\pi}\\\frac{dA_2}{4\pi}\\1
    \end{pmatrix}.
\end{equation}
The similarity transformation is 
\begin{multline}
   \left(
\begin{array}{cccccc}
  & \ldots  &  & 0 & 0 & \frac{j}{2} \\
 \ldots  & K_{IJ} & \ldots  & 0 & 0 & 0 \\
 & \ldots  &  & t_1 & t_2 & 0\\
 0 & 0 & t_1 & m_1 & n & 0 \\
 0 & 0 & t_2 & n & m_2 & 0\\
 \frac{j}{2} & 0 & 0 & 0 & 0 & 0 \\
\end{array}
\right) \\\simeq \left(
\begin{array}{cccccc}
  & \ldots  &  & W_{N,1} & 0 & W_{1,1}\frac{j}{2}  \\
 \ldots  & W^\intercal K_{IJ} W & \ldots  & \cdots & \cdots & 0 \\
 & \ldots  &  & W_{N,N}t_1 & W_{N,N}t_2 & W_{1,N}\frac{j}{2} \\
 W_{N,1} & \cdots & W_{N,N}t_1 & m_1 & n & 0 \\
 0 & 0 & W_{N,N}t_2 & n & m_2 & 0\\
 W_{1,1}\frac{j}{2} & \cdots & W_{1,N}\frac{j}{2} & 0 & 0 & 0 \\
\end{array}
\right) .
\end{multline}

\section{Anomaly in $U(1)$ and $U(1)\times U(1)$ Maxwell theory}\label{app:u1anomaly}
\paragraph{Review of $\bZ_{12}$ anomaly of $U(1)$ Maxwell theory}
For single $U(1)$ the mapping class group is isomorphic to $\SL(2,\bZ)$. The $\SL(2,\bZ)$ symmetry of the 3+1d Maxwell theory can act projectively on its 2+1d boundary, namely, the lifted $\hat{S}$ and $\hat{T}$ satisfy the group relations only up to stacking the invertible phase. In 2+1d, the generator of the invertible phases is $p+ip$ superconductor with chiral central charge $1/2$. Two copies of $p+ip$ are equivalent to $U(1)_1$. Denoting the stacking of a layer of $p+ip$ as operator $X$,
\begin{equation}
    \hat{S}^4 = X^p,\quad (\hat{S}\hat{T})^3 = X^q.
\end{equation}
If we redefine $\hat{S}'=X^a\hat{S},\; \hat{T}'=X^b\hat{T}$, then,
\begin{equation}
    p\rightarrow p+4a,\; q\rightarrow q+3(a+b),
\end{equation}
which gives the equivalence relation. Therefore, the different classes of projective action is given by,
\begin{equation}
    \frac{\bZ^2}{\langle (4,3),(0,3)\rangle} \cong \bZ_{12}
\end{equation}
which is consistent with \cite{Seiberg:2018ntt}. From the action of $(ST)^3$ on the $U(1)$ symmetric theory,
\begin{equation}
    (ST)^3: \cL[A] \rightarrow \cL[A]+\frac{1}{4\pi}ada, \; S^4: \cL[A] \rightarrow \cL[A] 
\end{equation}
we have,
\begin{equation}
    S^4=1,\; (ST)^3=X^2=U(1)_1.
\end{equation}
To match the common notation, we redefine $S'=X^2 S,\; T'=T$,
\begin{equation}
    S'^4= (S' T')^3 =X^8 
\end{equation}
which recovers the anomaly index $\nu=8\in \bZ_{12}$ of the Maxwell theory as the common convention $\hat{S}^4 =(\hat{S}\hat{T})^3 = X^n$ with $n\in \bZ_{12}$.

\paragraph{Derivation of $g_5^5$}
Recall that,
\begin{equation}
    g_5=T R_2 R_1 S:\;\cL[A_1,A_2]\rightarrow \cL[a,A_1]+\frac{1}{4\pi}A_1dA_1+\frac{1}{2\pi}ad(A_1+A_2)
\end{equation}
and we define the following notations for convenience,
\begin{equation}
    CS(x)=\frac{1}{4\pi}xdx,\; BF(x,y)=\frac{1}{2\pi}xdy
\end{equation}
Let $x_{-1}=B,x_0=A$,
\begin{equation}
    g_5^5: Z[A,B]\rightarrow \int\prod_{i=1}^5\cD x_i Z[x_5,x_4]e^{\i\sum_{j=0}^4(CS(x_j)+BF(x_{j+1},x_{j}+x_{j-1}))}
\end{equation}
Define $u=x_1+x_2+x_3$,
\begin{align}
    &CS(A)+CS(u)+CS(x_4)+BF(x_4,u)+BF(x_5,x_4+u) \nonumber\\
    &+BF(x_1,A+B-x_4-x_5)+BF(x_2,A-x_5) \\
    = &CS(A)+CS(u)+CS(x_4)+BF(x_4,u)+BF(A,x_4+u)+BF(x_1,B-x_4) \\
    = &CS(A)+CS(u)+CS(B)+BF(B,u)+BF(A,B+u)\\
    =&CS(u+A+B)
\end{align}
where we integrate over $x_2,x_1$ for the first and second equal signs. Therefore,
\begin{equation}
     g_5^5: \cL[A,B]\rightarrow \cL[A,B]+CS(u+A+B)
\end{equation}
where $\cL[A,B]+CS(u+A+B)$ is the decoupled $U(1)_1$.

\paragraph{Details on the $\bZ_{10}$ anomaly for $U(1)^2$} As pointed out in \cite{Seiberg:2018ntt}, the electromagnetic duality for $U(1)^2$ is captured by $H^2(B\Gamma_2,\bZ) \cong \bZ_{10}$, where $\Gamma_2$ is the mapping class group of genus 2 manifold. There is a symplectic representation of $\Gamma_2$, $\rho:\Gamma_2\rightarrow \Sp(4,\bZ)$ \cite{farb2011primer}, and $\rho$ is surjective, meaning $\rho(h)=g$ doesn't give a canonical $h$. $\rho$ of the 5 generators of $\Gamma_2$ is given by,
\begin{equation}
    \rho(\tau_1) = T, \; \rho(\tau_2) = S TS^{-1}, \; \rho(\tau_3) = R_2^{-1} R_1 T R_1 R_2, \; \rho(\tau_4) = R_1 S T S^{-1} R_1, \;   \rho(\tau_5) = R_1TR_1.
\end{equation}
Define $F=\tau_1\tau_2\tau_3\tau_4$, $F^5 =\iota$, where $\iota$ is the genus-two hyperelliptic involution \cite{farb2011primer}. $F^{10}=1$, and therefore, $F^2$ is order 5. Using the representation given above, our order-5 duality matrix $\sg_5$ is conjugated equal to,
\begin{equation}
    \sg_5 = \sh \rho(F)^2 \sh^{-1} = \rho(H F^2 H^{-1})
\end{equation}
where $\sh=ST^{-1}R_2R_1$, and choose $\sh=\rho(\tau_3\tau_2\tau_4 \tau_3\tau_5 \tau_4) \equiv \rho(H)$.
Then we can define an element $H_5 = H F^2 H^{-1} \in \Gamma_2$, whose representation $\rho(H_5) = \sg_5$. Therefore, we find the lift of $\sg_5$ to $H_5 \in\Gamma_2$.

$H^2(B\Gamma_2,\bZ) \cong \bZ_{10}$ basically describes the 1d unitary irrep of the abelianzation of $\Gamma_2$, as $ H^2(B\Gamma_2,\bZ) \cong H^1(B\Gamma_2,U(1)) \cong \Hom(\Gamma_2,U(1))$. We take $u\in \bZ_{10}$ and let $\chi_u(\tau_i)=e^{2\pi i /10}$, hence, $\chi_u(H_5)=e^{8\times2\pi i /10}$, then $H_5$ corresponds to $8\in \bZ_{10}$.

The $\bZ_5$ subgroup is the largest prime order subgroup of $\Sp(4,\bZ)$. When restricting to the $\bZ_5$ subgroup, $i: \langle H_5 \rangle = \bZ_{5} \hookrightarrow \Gamma_2$, let $y\in H^2(B\bZ_5,\bZ)$ with $\chi_y(H_5)=e^{2\pi i/5}$. Since $\chi_u(H_5)=e^{8\times2\pi i /10}=e^{4\times 2\pi i/5}$, then $i^* u = 4y$. Therefore, $i^*(8u)=32y=2y \mod 5$, so the anomaly is $2\in \bZ_5$ is consistent with the anomaly of the Maxwell theory being $8$.

\paragraph{Analogous $\bZ_2\times \bZ_2$ action of $g_5$}
There is an analogous $g_5$ action in $2k$-d quantum field theory with $k-1$-form $\bZ_2\times \bZ_2$ symmetry. Following \cite{Bhardwaj:2020ymp}, the corresponding action is given by,
\begin{equation}
    \g_5: Z[B_1,B_2]\rightarrow \frac{Q[B_1]}{\sqrt{\cN}}\sum_{b}(-1)^{\langle b\cup (B_1+B_2) \rangle}Z[b,B_1]
\end{equation}
where $1/\sqrt{\cN}$ is the normalization factor, $B_i$ and $b$ are the background and dynamical $\bZ_2$ $k$-cochain, $(-1)^{\langle \cdot,\cdot\rangle}$ is the intersection pairing, and $Q[B]= i^{q(B)}$, where $q(B)$ is a $\bZ_4$-valued function, the so-called quadratic refinement of the intersection pairing,
\begin{equation}
    q(B+B')= q(B)+q(B')+2{\langle B,B'\rangle}
\end{equation}
$\g_5$ is then the analogous twisted gauging, i.e. gauging and stacking fermionic SPT (in higher dimension, it corresponds to quadratic-refinement invertible phase) \cite{Karch:2019lnn,Gaiotto:2020iye,Lu:2024ytl,Huang:2024ror}. Similar to the derivation in the $U(1)^2$ case and \cite{Bhardwaj:2020ymp}, one can find,
\begin{equation}
    \g_5^5 : Z[B_1,B_2]\rightarrow \left(\frac{1}{\sqrt{\cN}}\sum_{b'}Q[b']\right)Z[B_1,B_2]
\end{equation}
where $\frac{1}{\sqrt{\cN}}\sum_{b'}Q[b']$ is the Brown-Kervaire invariant of the quadratic refinement and it is $\bZ_8$-valued and signifies the anomaly on $\Pin^-$ manifold. In 2d, it corresponds to the time-reversal invariant Kitaev chain \cite{Fidkowski:2009dba,Kapustin:2014dxa}. 

\bibliographystyle{JHEP}
\bibliography{biblio.bib}

@article{Karch:2019lnn,
    author = "Karch, Andreas and Tong, David and Turner, Carl",
    title = "{A Web of 2d Dualities: ${\bf Z}_2$ Gauge Fields and Arf Invariants}",
    eprint = "1902.05550",
    archivePrefix = "arXiv",
    primaryClass = "hep-th",
    doi = "10.21468/SciPostPhys.7.1.007",
    journal = "SciPost Phys.",
    volume = "7",
    pages = "007",
    year = "2019"
}

@article{Turner:2019wnh,
    author = "Turner, Carl",
    title = "{Dualities in 2+1 Dimensions}",
    eprint = "1905.12656",
    archivePrefix = "arXiv",
    primaryClass = "hep-th",
    doi = "10.22323/1.349.0001",
    journal = "PoS",
    volume = "Modave2018",
    pages = "001",
    year = "2019"
}

@article{Karch:2016sxi,
    author = "Karch, Andreas and Tong, David",
    title = "{Particle-Vortex Duality from 3d Bosonization}",
    eprint = "1606.01893",
    archivePrefix = "arXiv",
    primaryClass = "hep-th",
    doi = "10.1103/PhysRevX.6.031043",
    journal = "Phys. Rev. X",
    volume = "6",
    number = "3",
    pages = "031043",
    year = "2016"
}

@article{Seiberg:2016gmd,
    author = "Seiberg, Nathan and Senthil, T. and Wang, Chong and Witten, Edward",
    title = "{A Duality Web in 2+1 Dimensions and Condensed Matter Physics}",
    eprint = "1606.01989",
    archivePrefix = "arXiv",
    primaryClass = "hep-th",
    doi = "10.1016/j.aop.2016.08.007",
    journal = "Annals Phys.",
    volume = "374",
    pages = "395--433",
    year = "2016"
}

@article{Son:2015xqa,
    author = "Son, Dam Thanh",
    title = "{Is the Composite Fermion a Dirac Particle?}",
    eprint = "1502.03446",
    archivePrefix = "arXiv",
    primaryClass = "cond-mat.mes-hall",
    reportNumber = "EFI-15-6",
    doi = "10.1103/PhysRevX.5.031027",
    journal = "Phys. Rev. X",
    volume = "5",
    number = "3",
    pages = "031027",
    year = "2015"
}

@article{Senthil:2018cru,
    author = "Senthil, T. and Son, Dam Thanh and Wang, Chong and Xu, Cenke",
    title = "{Duality between $(2+1)d$ Quantum Critical Points}",
    eprint = "1810.05174",
    archivePrefix = "arXiv",
    primaryClass = "cond-mat.str-el",
    doi = "10.1016/j.physrep.2019.09.001",
    journal = "Phys. Rept.",
    volume = "827",
    pages = "1--48",
    year = "2019"
}

@article{Karch:2025hsj,
    author = "Karch, Andreas and Spieler, Ryan C.",
    title = "{Critical theories connecting gapped phases with {\ensuremath{\mathbb{Z}}}$_{2}$ {\texttimes} {\ensuremath{\mathbb{Z}}}$_{2}$ symmetry from the duality web}",
    eprint = "2502.14032",
    archivePrefix = "arXiv",
    primaryClass = "cond-mat.str-el",
    doi = "10.1007/JHEP06(2025)028",
    journal = "JHEP",
    volume = "06",
    pages = "028",
    year = "2025"
}

@misc{Tong:2016kpv,
    author = "Tong, David",
    title = "{Lectures on the Quantum Hall Effect}",
    eprint = "1606.06687",
    archivePrefix = "arXiv",
    primaryClass = "hep-th",
    note = "{TIFR lectures}",
    month = "6",
    year = "2016"
}

@article{Lu:2013jqa,
    author = "Lu, Yuan-Ming and Vishwanath, Ashvin",
    title = "{Classification and properties of symmetry-enriched topological phases: Chern-Simons approach with applications to Z$_2$ spin liquids}",
    eprint = "1302.2634",
    archivePrefix = "arXiv",
    primaryClass = "cond-mat.str-el",
    doi = "10.1103/PhysRevB.93.155121",
    journal = "Phys. Rev. B",
    volume = "93",
    number = "15",
    pages = "155121",
    year = "2016"
}

@article{Lu:2012dt,
    author = "Lu, Yuan-Ming and Vishwanath, Ashvin",
    title = "{Theory and classification of interacting 'integer' topological phases in two dimensions: A Chern-Simons approach}",
    eprint = "1205.3156",
    archivePrefix = "arXiv",
    primaryClass = "cond-mat.str-el",
    doi = "10.1103/PhysRevB.86.125119",
    journal = "Phys. Rev. B",
    volume = "86",
    number = "12",
    pages = "125119",
    year = "2012",
    note = "[Erratum: Phys.Rev.B 89, 199903 (2014)]"
}

@article{Seiberg:2018ntt,
    author = "Seiberg, Nathan and Tachikawa, Yuji and Yonekura, Kazuya",
    title = "{Anomalies of Duality Groups and Extended Conformal Manifolds}",
    eprint = "1803.07366",
    archivePrefix = "arXiv",
    primaryClass = "hep-th",
    reportNumber = "IPMU-18-0044",
    doi = "10.1093/ptep/pty069",
    journal = "PTEP",
    volume = "2018",
    number = "7",
    pages = "073B04",
    year = "2018"
}

@article{Bhardwaj:2020ymp,
    author = "Bhardwaj, Lakshya and Lee, Yasunori and Tachikawa, Yuji",
    title = "{$SL(2,\mathbb{Z})$ action on QFTs with $\mathbb{Z}_2$ symmetry and the Brown-Kervaire invariants}",
    eprint = "2009.10099",
    archivePrefix = "arXiv",
    primaryClass = "hep-th",
    reportNumber = "IPMU-20-0101",
    doi = "10.1007/JHEP11(2020)141",
    journal = "JHEP",
    volume = "11",
    pages = "141",
    year = "2020"
}

@article{DiPietro:2019hqe,
    author = "Di Pietro, Lorenzo and Gaiotto, Davide and Lauria, Edoardo and Wu, Jingxiang",
    title = "{3d Abelian Gauge Theories at the Boundary}",
    eprint = "1902.09567",
    archivePrefix = "arXiv",
    primaryClass = "hep-th",
    doi = "10.1007/JHEP05(2019)091",
    journal = "JHEP",
    volume = "05",
    pages = "091",
    year = "2019"
}

@article{Dimofte:2011ju,
    author = "Dimofte, Tudor and Gaiotto, Davide and Gukov, Sergei",
    title = "{Gauge Theories Labelled by Three-Manifolds}",
    eprint = "1108.4389",
    archivePrefix = "arXiv",
    primaryClass = "hep-th",
    reportNumber = "CALT-68-2847",
    doi = "10.1007/s00220-013-1863-2",
    journal = "Commun. Math. Phys.",
    volume = "325",
    pages = "367--419",
    year = "2014"
}

@article{hua1949generators,
  title={On the generators of the symplectic modular group},
  author={Hua, LK and Reiner, Irving},
  journal={Transactions of the American Mathematical Society},
  volume={65},
  number={3},
  pages={415--426},
  year={1949},
  publisher={JSTOR}
}

@article{scarola2001phase,
  title={Phase diagram of bilayer composite fermion states},
  author={Scarola, VW and Jain, JK},
  journal={Physical Review B},
  volume={64},
  number={8},
  pages={085313},
  year={2001},
  publisher={APS}
}

@article{Nguyen:2024jip,
    author = "Nguyen, Ron Q. and Zhang, Naiyuan J. and Khurana-Batra, Navketan and Alkidim, Sarah and Liu, Xiaoxue and Watanabe, Kenji and Taniguchi, Takashi and Feldman, D. E. and Li, J. I. A.",
    title = "{Bilayer Excitons in the Laughlin Fractional Quantum Hall State}",
    eprint = "2410.24208",
    archivePrefix = "arXiv",
    primaryClass = "cond-mat.mes-hall",
    month = "10",
    year = "2024"
}

@inproceedings{Witten:2003ya,
    author = "Witten, Edward",
    title = "{SL(2,Z) action on three-dimensional conformal field theories with Abelian symmetry}",
    booktitle = "{From Fields to Strings: Circumnavigating Theoretical Physics: A Conference in Tribute to Ian Kogan}",
    eprint = "hep-th/0307041",
    archivePrefix = "arXiv",
    pages = "1173--1200",
    month = "7",
    year = "2003"
}

@article{Kitaev:2005hzj,
    author = "Kitaev, Alexei",
    title = "{Anyons in an exactly solved model and beyond}",
    eprint = "cond-mat/0506438",
    archivePrefix = "arXiv",
    doi = "10.1016/j.aop.2005.10.005",
    journal = "Annals Phys.",
    volume = "321",
    number = "1",
    pages = "2--111",
    year = "2006"
}

@article{Wen:1992uk,
    author = "Wen, X. G. and Zee, A.",
    title = "{A Classification of Abelian quantum Hall states and matrix formulation of topological fluids}",
    reportNumber = "NSF-ITP-92-10",
    doi = "10.1103/PhysRevB.46.2290",
    journal = "Phys. Rev. B",
    volume = "46",
    pages = "2290--2301",
    year = "1992"
}

@article{Antinucci:2024zjp,
    author = "Antinucci, Andrea and Benini, Francesco",
    title = "{Anomalies and gauging of U(1) symmetries}",
    eprint = "2401.10165",
    archivePrefix = "arXiv",
    primaryClass = "hep-th",
    reportNumber = "SISSA 01/2024/FISI",
    doi = "10.1103/PhysRevB.111.024110",
    journal = "Phys. Rev. B",
    volume = "111",
    number = "2",
    pages = "024110",
    year = "2025"
}

@article{Seiberg:2016rsg,
    author = "Seiberg, Nathan and Witten, Edward",
    title = "{Gapped Boundary Phases of Topological Insulators via Weak Coupling}",
    eprint = "1602.04251",
    archivePrefix = "arXiv",
    primaryClass = "cond-mat.str-el",
    doi = "10.1093/ptep/ptw083",
    journal = "PTEP",
    volume = "2016",
    number = "12",
    pages = "12C101",
    year = "2016"
}

@article{barkeshliSymmetryFractionalizationDefects2019a,
    title = {Symmetry {Fractionalization}, {Defects}, and {Gauging} of {Topological} {Phases}},
    volume = {100},
    issn = {2469-9950, 2469-9969},
    url = {http://arxiv.org/abs/1410.4540},
    doi = {10.1103/PhysRevB.100.115147},
    number = {11},
    urldate = {2024-01-22},
    journal = {Physical Review B},
    author = {Barkeshli, Maissam and Bonderson, Parsa and Cheng, Meng and Wang, Zhenghan},
    month = sep,
    year = {2019},
    note = {arXiv:1410.4540 [cond-mat, physics:hep-th, physics:math-ph, physics:quant-ph]},
    pages = {115147},
}

@article{Hung:2013nla,
    author = "Hung, Ling-Yan and Wan, Yidun",
    title = "{K matrix Construction of Symmetry-Enriched Phases of Matter}",
    eprint = "1302.2951",
    archivePrefix = "arXiv",
    primaryClass = "cond-mat.str-el",
    doi = "10.1103/PhysRevB.87.195103",
    journal = "Phys. Rev. B",
    volume = "87",
    number = "19",
    pages = "195103",
    year = "2013"
}

@article{Cheng:2022nds,
    author = "Cheng, Meng and Jian, Chao-Ming",
    title = "{Gauging U(1) symmetry in (2+1)d topological phases}",
    eprint = "2201.07239",
    archivePrefix = "arXiv",
    primaryClass = "cond-mat.str-el",
    doi = "10.21468/SciPostPhys.12.6.202",
    journal = "SciPost Phys.",
    volume = "12",
    number = "6",
    pages = "202",
    year = "2022"
}

@article{Kohn:1961zz,
  author       = {Kohn, Walter},
  title        = {Cyclotron Resonance and de Haas-van Alphen Oscillations of an Interacting Electron Gas},
  journal      = {Phys. Rev.},
  volume       = {123},
  pages        = {1242--1244},
  year         = {1961},
  doi          = {10.1103/PhysRev.123.1242}
}

@article{Hoyos:2011ez,
  author       = {Hoyos, Carlos and Son, Dam Thanh},
  title        = {Hall Viscosity and Electromagnetic Response},
  journal      = {Phys. Rev. Lett.},
  volume       = {108},
  pages        = {066805},
  year         = {2012},
  doi          = {10.1103/PhysRevLett.108.066805},
  eprint       = {1109.2651},
  archivePrefix = {arXiv},
  primaryClass = {cond-mat.mes-hall}
}

@article{Haldane:1983xm,
  author       = {Haldane, F. D. M.},
  title        = {Fractional Quantization of the Hall Effect: A Hierarchy of Incompressible Quantum Fluid States},
  journal      = {Phys. Rev. Lett.},
  volume       = {51},
  number       = {7},
  pages        = {605--608},
  year         = {1983},
  doi          = {10.1103/PhysRevLett.51.605}
}

@article{Halperin:1983mmn,
  author       = {Halperin, B. I.},
  title        = {Theory of the Quantized Hall Conductance},
  journal      = {Helv. Phys. Acta},
  volume       = {56},
  pages        = {75--102},
  year         = {1983}
}

@article{Halperin:1984zz,
  author       = {Halperin, B. I.},
  title        = {Statistics of Quasiparticles and the Hierarchy of Fractional Quantized Hall States},
  journal      = {Phys. Rev. Lett.},
  volume       = {52},
  number       = {18},
  pages        = {1583--1586},
  year         = {1984},
  doi          = {10.1103/PhysRevLett.52.1583},
  note         = {Erratum: Phys. Rev. Lett. 52, 2390 (1984), doi:10.1103/PhysRevLett.52.2390.4}
}

@article{Jain:1989zz,
  author       = {Jain, J. K.},
  title        = {Composite-Fermion Approach for the Fractional Quantum Hall Effect},
  journal      = {Phys. Rev. Lett.},
  volume       = {63},
  number       = {2},
  pages        = {199--202},
  year         = {1989},
  doi          = {10.1103/PhysRevLett.63.199}
}

@article{Blok:1990mc,
  author       = {Blok, B. and Wen, X. G.},
  title        = {Effective Theories of the Fractional Quantum Hall Effect: Hierarchy Construction},
  journal      = {Phys. Rev. B},
  volume       = {42},
  number       = {13},
  pages        = {8145--8156},
  year         = {1990},
  doi          = {10.1103/PhysRevB.42.8145}
}

@article{Read:1990zza,
  author       = {Read, N.},
  title        = {Excitation Structure of the Hierarchy Scheme in the Fractional Quantum Hall Effect},
  journal      = {Phys. Rev. Lett.},
  volume       = {65},
  number       = {12},
  pages        = {1502--1505},
  year         = {1990},
  doi          = {10.1103/PhysRevLett.65.1502}
}

@article{Wen:1995qn,
  author       = {Wen, Xiao-Gang},
  title        = {Topological Orders and Edge Excitations in Fractional Quantum Hall States},
  journal      = {Adv. Phys.},
  volume       = {44},
  number       = {5},
  pages        = {405--473},
  year         = {1995},
  doi          = {10.1080/00018739500101566},
  eprint       = {cond-mat/9506066},
  archivePrefix = {arXiv}
}

@article{Barkeshli:2019vtb,
  author       = {Barkeshli, Maissam and Bonderson, Parsa and Cheng, Meng and Wang, Zhenghan},
  title        = {Symmetry Fractionalization, Defects, and Gauging of Topological Phases},
  journal      = {Phys. Rev. B},
  volume       = {100},
  number       = {11},
  pages        = {115147},
  year         = {2019},
  doi          = {10.1103/PhysRevB.100.115147},
  eprint       = {1410.4540},
  archivePrefix = {arXiv},
  primaryClass = {cond-mat.str-el}
}

@article{Cheng:2022jbr,
  author       = {Cheng, Meng and Hsin, Po-Shen and Jian, Chao-Ming},
  title        = {Gauging Lie Group Symmetry in {(2+1)d} Topological Phases},
  journal      = {SciPost Phys.},
  volume       = {14},
  number       = {5},
  pages        = {100},
  year         = {2023},
  doi          = {10.21468/SciPostPhys.14.5.100},
  eprint       = {2205.15347},
  archivePrefix = {arXiv},
  primaryClass = {cond-mat.str-el}
}

@article{Zhang:2024bye,
    author = "Zhang, Carolyn and Vishwanath, Ashvin and Wen, Xiao-Gang",
    title = "{Hierarchy construction for non-Abelian fractional quantum Hall states via anyon condensation}",
    eprint = "2406.12068",
    archivePrefix = "arXiv",
    primaryClass = "cond-mat.str-el",
    doi = "10.1103/jndb-435f",
    journal = "Phys. Rev. B",
    volume = "112",
    number = "12",
    pages = "125116",
    year = "2025"
}

@article{Spieler:2025hyr,
    author = "Spieler, Ryan C.",
    title = "{Exploring two dimensional ${\mathbb{Z}}_{2}$ invariant phases with time reversal symmetry and their transitions with topological operations}",
    eprint = "2504.20021",
    archivePrefix = "arXiv",
    primaryClass = "cond-mat.str-el",
    doi = "10.1007/JHEP10(2025)069",
    journal = "JHEP",
    volume = "10",
    pages = "069",
    year = "2025"
}

@article{Behr1975,
author = {Behr, Helmut},
journal = {Mathematische Zeitschrift},
pages = {47-56},
title = {Eine endliche Präsentation der symplektischen Gruppe Sp4 (Z).},
url = {http://eudml.org/doc/172154},
volume = {141},
year = {1975},
}

@ARTICLE{2017jain-nonabelian,
       author = {{Hutasoit}, Jimmy A. and {Balram}, Ajit C. and {Mukherjee}, Sutirtha and {Wu}, Ying-Hai and {Mandal}, Sudhansu S. and {W{\'o}js}, A. and {Cheianov}, Vadim and {Jain}, J.~K.},
        title = "{The enigma of the {\ensuremath{\nu}} =2 +3 /8 fractional quantum Hall effect}",
      journal = {\prb},
         year = 2017,
        month = mar,
       volume = {95},
       number = {12},
          eid = {125302},
        pages = {125302},
          doi = {10.1103/PhysRevB.95.125302},
archivePrefix = {arXiv},
       eprint = {1605.07324},
 primaryClass = {cond-mat.mes-hall},
       adsurl = {https://ui.adsabs.harvard.edu/abs/2017PhRvB..95l5302H}
}

@article{Wang:2020nmz,
    author = "Wang, Liang and Wang, Zhenghan",
    title = "{In and around Abelian anyon models}",
    eprint = "2004.12048",
    archivePrefix = "arXiv",
    primaryClass = "math.QA",
    doi = "10.1088/1751-8121/abc6c0",
    journal = "J. Phys. A",
    volume = "53",
    number = "50",
    pages = "505203",
    year = "2020"
}

@article{MooreRead1991,
    author  = {Moore, Gregory and Read, Nicholas},
    title   = {Nonabelions in the Fractional Quantum Hall Effect},
    journal = {Nucl. Phys. B},
    volume  = {360},
    number  = {2-3},
    pages   = {362--396},
    year    = {1991},
    doi     = {10.1016/0550-3213(91)90407-O}
}

@article{ReadRezayi1999,
    author        = {Read, Nicholas and Rezayi, E.},
    title         = {Beyond Paired Quantum Hall States: Parafermions and
                     Incompressible States in the First Excited Landau Level},
    journal       = {Phys. Rev. B},
    volume        = {59},
    pages         = {8084--8092},
    year          = {1999},
    doi           = {10.1103/PhysRevB.59.8084},
    eprint        = {cond-mat/9809384},
    archivePrefix = {arXiv}
}

@article{Kong2014,
    author        = {Kong, Liang},
    title         = {Anyon Condensation and Tensor Categories},
    journal       = {Nucl. Phys. B},
    volume        = {886},
    pages         = {436--482},
    year          = {2014},
    doi           = {10.1016/j.nuclphysb.2014.07.003},
    eprint        = {1307.8244},
    archivePrefix = {arXiv},
    primaryClass  = {cond-mat.str-el}
}

@article{BondersonSlingerland2008,
    author        = {Bonderson, Parsa and Slingerland, J. K.},
    title         = {Fractional Quantum Hall Hierarchy and the Second
                     Landau Level},
    journal       = {Phys. Rev. B},
    volume        = {78},
    pages         = {125323},
    year          = {2008},
    doi           = {10.1103/PhysRevB.78.125323},
    eprint        = {0711.3204},
    archivePrefix = {arXiv},
    primaryClass  = {cond-mat.mes-hall}
}

@article{BarkeshliWen2010TwoComponent,
    author        = {Barkeshli, Maissam and Wen, Xiao-Gang},
    title         = {Non-Abelian Two-Component Fractional Quantum Hall States},
    journal       = {Phys. Rev. B},
    volume        = {82},
    pages         = {233301},
    year          = {2010},
    doi           = {10.1103/PhysRevB.82.233301},
    eprint        = {0906.0356},
    archivePrefix = {arXiv},
    primaryClass  = {cond-mat.mes-hall}
}

@article{BarkeshliWen2011Bilayer,
    author        = {Barkeshli, Maissam and Wen, Xiao-Gang},
    title         = {Bilayer Quantum Hall Phase Transitions and the Orbifold
                     Non-Abelian Fractional Quantum Hall States},
    journal       = {Phys. Rev. B},
    volume        = {84},
    pages         = {115121},
    year          = {2011},
    doi           = {10.1103/PhysRevB.84.115121},
    eprint        = {1010.4270},
    archivePrefix = {arXiv},
    primaryClass  = {cond-mat.str-el}
}

@article{benson2018cohomology,
  title={Cohomology of symplectic groups and Meyer’s signature theorem},
  author={Benson, Dave and Campagnolo, Caterina and Ranicki, Andrew and Rovi, Carmen},
  journal={Algebraic \& Geometric Topology},
  volume={18},
  number={7},
  pages={4069--4091},
  year={2018},
  eprint = "1710.04851",
  archivePrefix = "arXiv",
  publisher={Mathematical Sciences Publishers}
}

@article{Wang:2017txt,
    author = "Wang, Chong and Nahum, Adam and Metlitski, Max A. and Xu, Cenke and Senthil, T.",
    title = "{Deconfined quantum critical points: symmetries and dualities}",
    eprint = "1703.02426",
    archivePrefix = "arXiv",
    primaryClass = "cond-mat.str-el",
    doi = "10.1103/PhysRevX.7.031051",
    journal = "Phys. Rev. X",
    volume = "7",
    number = "3",
    pages = "031051",
    year = "2017"
}

@article{Xu:2015lxa,
    author = "Xu, Cenke and You, Yi-Zhuang",
    title = "{Self-dual Quantum Electrodynamics as Boundary State of the three dimensional Bosonic Topological Insulator}",
    eprint = "1510.06032",
    archivePrefix = "arXiv",
    primaryClass = "cond-mat.str-el",
    doi = "10.1103/PhysRevB.92.220416",
    journal = "Phys. Rev. B",
    volume = "92",
    number = "22",
    pages = "220416",
    year = "2015"
}

@article{Cheng:2016pdn,
    author = "Cheng, Meng and Xu, Cenke",
    title = "{Series of (2+1)-dimensional stable self-dual interacting conformal field theories}",
    eprint = "1609.02560",
    archivePrefix = "arXiv",
    primaryClass = "cond-mat.str-el",
    doi = "10.1103/PhysRevB.94.214415",
    journal = "Phys. Rev. B",
    volume = "94",
    number = "21",
    pages = "214415",
    year = "2016"
}

@article{Jian:2017chw,
    author = "Jian, Chao-Ming and Rasmussen, Alex and You, Yi-Zhuang and Xu, Cenke",
    title = "{Emergent Symmetry and Tricritical Points near the deconfined Quantum Critical Point}",
    eprint = "1708.03050",
    archivePrefix = "arXiv",
    primaryClass = "cond-mat.str-el",
    month = "8",
    year = "2017"
}

@article{Lu:2021ucu,
    author = "Lu, Da-Chuan and Xu, Cenke and You, Yi-Zhuang",
    title = "{Self-duality protected multicriticality in deconfined quantum phase transitions}",
    eprint = "2104.05147",
    archivePrefix = "arXiv",
    primaryClass = "cond-mat.str-el",
    doi = "10.1103/PhysRevB.104.205142",
    journal = "Phys. Rev. B",
    volume = "104",
    number = "20",
    pages = "205142",
    year = "2021"
}

@article{Dumitrescu:2026vre,
    author = "Dumitrescu, Thomas T. and Niro, Pierluigi and Thorngren, Ryan",
    title = "{From QED$_3$ to Self-Dual Multicriticality in the Fradkin-Shenker Model}",
    eprint = "2602.23420",
    archivePrefix = "arXiv",
    primaryClass = "cond-mat.str-el",
    month = "2",
    year = "2026"
}

@article{Gaillard:1981rj,
    author = {Gaillard, Mary K. and Zumino, Bruno},
    title = {Duality Rotations for Interacting Fields},
    reportNumber = {LBL-12608, UCB-PTH-81/1},
    doi = {10.1016/0550-3213(81)90527-7},
    journal = {Nucl. Phys. B},
    volume = {193},
    pages = {221--244},
    year = {1981}
}

@article{Kapustin:2009av,
    author = {Kapustin, Anton and Tikhonov, Mikhail},
    title = {Abelian Duality, Walls and Boundary Conditions in Diverse Dimensions},
    eprint = {0904.0840},
    archivePrefix = {arXiv},
    primaryClass = {hep-th},
    doi = {10.1088/1126-6708/2009/11/006},
    journal = {JHEP},
    volume = {11},
    pages = {006},
    year = {2009}
}

@article{Harer1983,
    author = {Harer, John},
    title = {The Second Homology Group of the Mapping Class Group of an Orientable Surface},
    journal = {Invent. Math.},
    volume = {72},
    number = {2},
    pages = {221--239},
    year = {1983},
    doi = {10.1007/BF01389321}
}

@article{KorkmazStipsicz2003,
    author = {Korkmaz, Mustafa and Stipsicz, Andr{\'a}s I.},
    title = {The Second Homology Groups of Mapping Class Groups of Orientable Surfaces},
    journal = {Math. Proc. Cambridge Philos. Soc.},
    volume = {134},
    number = {3},
    pages = {479--489},
    year = {2003},
    doi = {10.1017/S0305004102006461}
}

@article{Watanabe:2020ngm,
    author = {Watanabe, Haruki},
    title = {Counting Rules of Nambu--Goldstone Modes},
    eprint = {1904.00569},
    archivePrefix = {arXiv},
    primaryClass = {cond-mat.other},
    doi = {10.1146/annurev-conmatphys-031119-050644},
    journal = {Annu. Rev. Condens. Matter Phys.},
    volume = {11},
    pages = {169--187},
    year = {2020}
}

@article{Jia:2026tfh,
    author = "Jia, Qiang and Zhang, Yi",
    title = "{Flat Gauging of Continuous (Non-invertible) Symmetries and Non-compact BF SymTFT for Compact Boson}",
    eprint = "2606.15732",
    archivePrefix = "arXiv",
    primaryClass = "hep-th",
    month = "6",
    year = "2026"
}

@article{Brennan:2024fgj,
    author = "Brennan, T. Daniel and Sun, Zhengdi",
    title = "{A SymTFT for continuous symmetries}",
    eprint = "2401.06128",
    archivePrefix = "arXiv",
    primaryClass = "hep-th",
    doi = "10.1007/JHEP12(2024)100",
    journal = "JHEP",
    volume = "12",
    pages = "100",
    year = "2024"
}

@article{Apruzzi:2024htg,
    author = "Apruzzi, Fabio and Bedogna, Francesco and Dondi, Nicola",
    title = "{SymTh for non-finite symmetries}",
    eprint = "2402.14813",
    archivePrefix = "arXiv",
    primaryClass = "hep-th",
    doi = "10.1007/JHEP04(2026)153",
    journal = "JHEP",
    volume = "04",
    pages = "153",
    year = "2026"
}

@article{Moradi:2022lqp,
    author = "Moradi, Heidar and Moosavian, Seyed Faroogh and Tiwari, Apoorv",
    title = "{Topological holography: Towards a unification of Landau and beyond-Landau physics}",
    eprint = "2207.10712",
    archivePrefix = "arXiv",
    primaryClass = "cond-mat.str-el",
    doi = "10.21468/SciPostPhysCore.6.4.066",
    journal = "SciPost Phys. Core",
    volume = "6",
    pages = "066",
    year = "2023"
}

@article{Huang:2023pyk,
    author = "Huang, Sheng-Jie and Cheng, Meng",
    title = "{Topological holography, quantum criticality, and boundary states}",
    eprint = "2310.16878",
    archivePrefix = "arXiv",
    primaryClass = "cond-mat.str-el",
    doi = "10.21468/SciPostPhys.18.6.213",
    journal = "SciPost Phys.",
    volume = "18",
    number = "6",
    pages = "213",
    year = "2025"
}

@book{farb2011primer,
  title={A primer on mapping class groups},
  author={Farb, Benson and Margalit, Dan},
  volume={49},
  year={2011},
  publisher={Princeton university press}
}

@article{balatsky1991singlet,
  title={Singlet quantum Hall effect and Chern-Simons theories},
  author={Balatsky, Alexander and Fradkin, Eduardo},
  journal={Physical Review B},
  volume={43},
  number={13},
  pages={10622},
  year={1991},
  publisher={APS}
}

@article{milovanovic1997invariant,
  title={Invariant structure of the hierarchy theory of fractional quantum Hall states with spin},
  author={Milovanovi{\'c}, M and Read, N},
  journal={Physical Review B},
  volume={56},
  number={3},
  pages={1461},
  year={1997},
  publisher={APS}
}

@article{Kapustin:2014dxa,
    author = "Kapustin, Anton and Thorngren, Ryan and Turzillo, Alex and Wang, Zitao",
    title = "{Fermionic Symmetry Protected Topological Phases and Cobordisms}",
    eprint = "1406.7329",
    archivePrefix = "arXiv",
    primaryClass = "cond-mat.str-el",
    doi = "10.1007/JHEP12(2015)052",
    journal = "JHEP",
    volume = "12",
    pages = "052",
    year = "2015"
}

@article{Fidkowski:2009dba,
    author = "Fidkowski, Lukasz and Kitaev, Alexei",
    title = "{The effects of interactions on the topological classification of free fermion systems}",
    eprint = "0904.2197",
    archivePrefix = "arXiv",
    primaryClass = "cond-mat.str-el",
    doi = "10.1103/PhysRevB.81.134509",
    journal = "Phys. Rev. B",
    volume = "81",
    pages = "134509",
    year = "2010"
}

@article{Gaiotto:2020iye,
    author = "Gaiotto, Davide and Kulp, Justin",
    title = "{Orbifold groupoids}",
    eprint = "2008.05960",
    archivePrefix = "arXiv",
    primaryClass = "hep-th",
    doi = "10.1007/JHEP02(2021)132",
    journal = "JHEP",
    volume = "02",
    pages = "132",
    year = "2021"
}

@article{Lu:2024ytl,
    author = "Lu, Da-Chuan and Sun, Zhengdi and You, Yi-Zhuang",
    title = "{Realizing triality and $p$-ality by lattice twisted gauging in (1+1)d quantum spin systems}",
    eprint = "2405.14939",
    archivePrefix = "arXiv",
    primaryClass = "cond-mat.str-el",
    doi = "10.21468/SciPostPhys.17.5.136",
    journal = "SciPost Phys.",
    volume = "17",
    number = "5",
    pages = "136",
    year = "2024"
}

@article{Huang:2024ror,
    author = "Huang, Sheng-Jie",
    title = "{Fermionic quantum criticality through the lens of topological holography}",
    eprint = "2405.09611",
    archivePrefix = "arXiv",
    primaryClass = "cond-mat.str-el",
    doi = "10.1103/PhysRevB.111.155130",
    journal = "Phys. Rev. B",
    volume = "111",
    number = "15",
    pages = "155130",
    year = "2025"
}

@article{Burgess:2000kj,
    author = "Burgess, C. P. and Dolan, Brian P.",
    title = "{Particle vortex duality and the modular group: Applications to the quantum Hall effect and other 2-D systems}",
    eprint = "hep-th/0010246",
    archivePrefix = "arXiv",
    reportNumber = "MCGILL-00-22, IASSNS-HEP-00-77",
    doi = "10.1103/PhysRevB.63.155309",
    journal = "Phys. Rev. B",
    volume = "63",
    pages = "155309",
    year = "2001"
}

@article{Burgess:2001sy,
    author        = "Burgess, C. P. and Dolan, Brian P.",
    title         = "{Duality and nonlinear response for quantum Hall systems}",
    eprint        = "cond-mat/0105621",
    archivePrefix = "arXiv",
    primaryClass  = "cond-mat.mes-hall",
    reportNumber  = "MCGILL-01-08, DIAS-STP-01-08",
    doi           = "10.1103/PhysRevB.65.155323",
    journal       = "Phys. Rev. B",
    volume        = "65",
    pages         = "155323",
    year          = "2002"
}

@article{Burgess:2007qa,
    author        = "Burgess, C. P. and Dolan, B. P.",
    title         = "{Modular symmetry, the semicircle law, and quantum Hall bilayers}",
    eprint        = "cond-mat/0701535",
    archivePrefix = "arXiv",
    primaryClass  = "cond-mat.mes-hall",
    reportNumber  = "DIAS-STP-06-25",
    doi           = "10.1103/PhysRevB.76.155310",
    journal       = "Phys. Rev. B",
    volume        = "76",
    pages         = "155310",
    year          = "2007"
}

@article{Lutken:2011zz,
    author        = {L{\"u}tken, C. A. and Ross, G. G.},
    title         = "{Experimental probes of emergent symmetries in the quantum Hall system}",
    eprint        = "1008.5257",
    archivePrefix = "arXiv",
    primaryClass  = "cond-mat.str-el",
    reportNumber  = "CERN-PH-TH-2010-192-OUTP-10-23P",
    doi           = "10.1016/j.nuclphysb.2011.04.020",
    journal       = "Nucl. Phys. B",
    volume        = "850",
    pages         = "321--338",
    year          = "2011"
}

@article{NissinenLutken2012,
    author        = {Nissinen, J. and L{\"u}tken, C. A.},
    title         = "{Renormalization-group potential for quantum Hall effects}",
    eprint        = "1111.4902",
    archivePrefix = "arXiv",
    primaryClass  = "cond-mat.str-el",
    doi           = "10.1103/PhysRevB.85.155123",
    journal       = "Phys. Rev. B",
    volume        = "85",
    number        = "15",
    pages         = "155123",
    year          = "2012"
}

@article{GeraedtsMotrunich2013,
    author        = "Geraedts, Scott D. and Motrunich, Olexei I.",
    title         = "{Exact realization of integer and fractional quantum Hall phases in $U(1)\times U(1)$ models in $(2+1)d$}",
    eprint        = "1302.1436",
    archivePrefix = "arXiv",
    primaryClass  = "cond-mat.str-el",
    doi           = "10.1016/j.aop.2013.03.017",
    journal       = "Annals Phys.",
    volume        = "334",
    pages         = "288--315",
    year          = "2013"
}

@article{Goldman:2018zfm,
    author        = "Goldman, Hart and Fradkin, Eduardo",
    title         = "{Loop Models, Modular Invariance, and Three Dimensional Bosonization}",
    eprint        = "1801.04936",
    archivePrefix = "arXiv",
    primaryClass  = "cond-mat.str-el",
    doi           = "10.1103/PhysRevB.97.195112",
    journal       = "Phys. Rev. B",
    volume        = "97",
    number        = "19",
    pages         = "195112",
    year          = "2018"
}

@article{burgisser1982elements,
  title={Elements of finite order in symplectic groups},
  author={B{\"u}rgisser, B},
  journal={Archiv der Mathematik},
  volume={39},
  number={6},
  pages={501--509},
  year={1982},
  publisher={Springer}
}

\end{document}